# Steps and terraces at quasicrystal surfaces. Application of the 6d-polyhedral model to the analysis of STM images of i-AlPdMn

L. BARBIER

DSM/DRECAM/SPCSI, CEA Saclay, 91191 Gif-sur-Yvette Cedex, France

and

D. GRATIAS

LEM CNRS-ONERA B.P. 72 92322 Châtillon cedex, FRANCE

6-d polyhedral models give a periodic description of aperiodic quasicrystals. There are powerful tools to describe their structural surface properties. Basis of the model for $F(2)$ icosahedral quasicrystals are given. This description is further used to interpret high resolution STM images of the $5f$ surface of i-AlPdMn which surface preparation was followed by He diffraction. It is found that both terrace structure and step-terrace height profiles in STM images can be consistently interpreted by the described model.

# I. Introduction

X-ray and neutron diffraction together with High Resolution Electron Microscopy (HREM) observations, allowed to explore the bulk structure of quasicrystals (QC) and several structural models have been proposed[1 2 3 4.] Description of these intriguing solids with 5-fold orientational symmetry (without translational symmetry) can be given by means of periodic $n$-d spaces ($n$>3, for instance $n$=6 for icosahedral AlPdMn alloys), giving within our physical 3-d space objects with the required symmetry properties. As for usual crystals, together with orientational and translational symmetries, the content of the $n$-d Bravais unit cell must be given, these defining the whole QC structure.

One crucial test for these $n$-d models, well fitting diffraction data, is the ability of these models to generate QCs configurations that can be compared to real space observations. For usual crystals, HREM observations[5,6] give real space images that are 2-d projection of the crystal structure which interpretation is not a so easy task. For QCs, similar interpretations of HREM images need a good description of the electron transmission through the aperiodic structure of a thin sample. Confirmation of some features of the proposed structural models were done in that way but remains somehow indirect.

In contrast, surface studies by Scanning Tunneling Microscopy (STM) allow getting real space configurations, without thickness averaging. Specific applications for surfaces of QC models have been proposed [7 8 9 10]. As for surfaces, the structure of extended terraces may be altered (segregation and/or surface reconstruction) with respect to the bulk. However, wide investigations of crystal surfaces of usual ordered alloy (like $Cu_3Au$ and Cu-Pd(17%) (cfc-$L1_2$ ordered[11 12]) or $Fe_3Al$[13] (bcc B2 or $DO_3$ ordered) have mainly shown that terraces exhibit a truncated bulk structure at the price of a step reorganization, in such a way that all terraces correspond to low energy planes. Terrace and step heights between wide terraces can be directly related to a bulk truncation. Such ideas appear to remain valid for QC[9 14] and make encouraging the exploration by Scanning Tunneling Microscopy (STM) of QC structures from observation of their surfaces. In return, beyond the investigation of one bulk QC structure, specific structural features of QC surfaces (nature of terraces, step heights, local structure, defects) have to be understood in view of their potential applications (reactivity, tribology etc ..).

If high resolution STM images have been early obtained (see[15], *d*-AlCoCu[16], *d*-AlNiCo[17], *i*-AlPdMn[18 19 20 21] *i*-AlCuFe[22]), it is only recently that high quality STM images at atomic resolution were obtained (i-AlPdMn[9 23 24 25 26]., i-AlCuFe[14 27]). Successful comparisons to tiling models (*i*-AlPdMn)[23 24 25 26], polyhedral model (*i*-AlPdMn[9]) or spherical model (*i*-AlCuFe[14]) were reached for some local surface configurations showing that the terrace structure indeed directly reflects the bulk structure. After these first successful attempts (see also[28] for a recent extended review), deeper investigations of QC models would allow interpreting step heights and distribution of main terraces giving a more complete view of the QCs surface structure.

The aim of the present contribution is to give an extended exploitation of the polyhedral $n$-d QC models with the aim of a better understanding of their surface structural properties. After a short introduction to the polyhedral model, the 6-d structure is described. Required for surface studies, properties of QC-directions and planes and their associated hyperplanes are obtained by application of the model. Application to the analysis of STM images of a well prepared QC surface ($5f$ i-AlPdMn), as shown by a He diffraction study, is further presented. For the experimentalist, detailed tools for application of the model to surface studies are explicitly given in the Appendix.

## II. Basis of quasicrystallography

### II.1 The 6-d models

Various approaches are able to describe 3d objects with icosaedral symmetry. Among them, tiling models are useful to decrypt the symmetry properties of aperiodic real space configurations of QCs. However, these geometrical models results from (and thence only contains) the specific symmetries of QCs, and cannot provide additional information on the internal structure of the unit cell. Knowing from diffraction data the QC symmetry, associated tiling models must apply and mapping by such models of observed QC configurations allows only checking that the expected symmetry properties are well observed in real space.

Polyhedral models are physical descriptions of the QC structure where nature and positions of atoms are defined from one unique unit cell embedded in a 6d-space (see below). In contrast with similar spherical models[2] (where random probabilities or harmonic expansion of the AS shape[10] must be introduced to avoid unphysical overlapping of atoms), polyhedral models have several great advantages: by construction the polyhedral shape of the 6-d Bravais unit cell avoid undesirable short distances and allows a deterministic description of the QC structure where 3-d wide QC configurations can be unambiguously generated. In the following, we focus on this type of model.

Due to their intrinsic aperiodicity, it is difficult to capture the main properties of QCs on the sole basis of their description in the 3-d real space. According to their properties of orientational symmetry, each diffraction peak (defined as an intense spot within the reciprocal space) can be indexed by a linear combination with *integer* coefficients of a set of $n$ unit vectors $\{|i\rangle\}$ where $n$ is greater than 3, the physical space dimension. In the following, the actual case of $i$-AlPdMn (or $i$-AlCuFe, both on a $F(2)$ lattice) will be considered where 6 vectors pointing on the vertices of an icosahedron are necessary to index the full set of diffraction peaks[1]. In that way, incommensurate coordinates are necessary in every basis of the 3-d real space to define points of the structure (in real or reciprocal space). Orientational symmetry makes the components indexing the QC lattice in real space a combination $(n+m\tau)$ of integers ($n$ and $m$) and one single irrational $\tau$ defining a $Z(\tau)$-module. For icosahedral symmetry, $\tau$ is the golden mean $\tau=\left(1+\sqrt{5}\right)/2$. Elements of $Z(\tau)$ on one 1-d axis can be also seen as the projection of points $(n,m)$ of a periodic 2-d lattice with orthonormal basis $\{z_1, z_\tau\}$, where $z_\tau$ makes an angle $atan(1/\tau)$ = 31.717° with the 1-d axis (as required the ratio of the length of the projected basis vectors is $\tau$). By extension of this periodic 2-d description of a 1-d aperiodic structure, a *periodic* description of a 3-d aperiodic QC is obtained within a 6-d hyperspace $E^6$ containing the $F(2)$ periodic cubic lattice, say $\Lambda$, with its associated natural orthonormal basis $\{|i\rangle\}$. This hyperspace can be further decomposed into the 3-d physical subspace $E_{//}$ with its associated orthonormal basis $\{|\alpha = x, y, z\rangle\}$and a 3-d subspace orthogonal to it $E_\perp$ (basis $\{|\beta = x_\perp, y_\perp, z_\perp\rangle\}$. Following the indexation proposed by Cahn, Shechtman and Gratias (CSG)[1], the basis vectors of $E_{//}$ and $E_\perp$, are chosen along three mutually perpendicular 2-fold directions. Each subspace being invariant through the Fm35 group, the projection matrices $\Pi^{//}$ (from $E^6$ to $E_{//}$) and $\Pi^\perp$ (from $E^6$ to $E_\perp$) must satisfy the commutation relations[29]:

$$\left[\Pi^{//}, g_e\right]=0 \qquad \left[\pi^{\perp}, g_e\right]=0$$

where $g_e$ are the generator elements of the group Fm35. These relations fully define the projection matrices and therefore the relative orientation of $E_{//}$ and $E_{\perp}$ in $E^6$. This makes the coordinates of the projection in $E_{//}$ of the basis vectors of $\Lambda$ reflecting the required incommensurability (i.e. giving 3-d coordinates that all belongs to the $Z(\tau)$-module, see the Appendix). It is also worth noting that the projections of $\{|i\rangle\}$ within $E_{//}$ and $E_{\perp}$ are along one of the six $5f$ directions of one regular icosahedron in $E_{//}$ and another in $E_{\perp}$, which are defined by the symmetry. The chosen projection scheme preserves a full symmetry between $E_{//}$ and $E_{\perp}$: any expression in one projection space transforms in the other by changing the numbers of the form $(u+v\tau)$ in $(v-u\tau)$. In addition, the unit length in $E_{\perp}$ being arbitrary, the same unit length in both projection spaces can be (without restriction) naturally chosen to preserve the cubic lattice in $E^6$ and orthogonality in $E_{//}$-$E_{\perp}$ maps. The unique length scale $a$ in $E^6$ ($E^6$ unit equal to the length of the hyper cubic primitive lattice parameter $\||i\rangle\|$) is 0.6451 nm for *i*-AlPdMn and 0.6314 nm for *i*-AlCuFe. All explicit length values in the present paper are given in $a$ unit.

The real bulk icosahedral structure is generated by a cut algorithm of $E^6$ by the 3-d physical space $E_{//}$. Like a point results from the intersection of a plane by a non parallel line, each point (atom) of the structure results from the cut of 3d volumes embedded in $E_{\perp}$, called Atomic Surfaces (ASs)[30] by the $E_{//}$ space. To get real configurations in our space $E_{//}$, the necessary projection operators from $E^6$ to $E_{//}$ or $E_{\perp}$ has to be written. The (3x6) projection matrices from $E^6$ ($\{|i\rangle\}$ basis) to $E_{//}$ or $E_{\perp}$ with the $\{|\alpha\rangle, |\beta\rangle\}$ basis are:

$$M^{//}_{\alpha,i} = \langle \alpha | i \rangle, \qquad M^{\perp}_{\beta,i} = \langle \beta | i \rangle,$$

or writing the projected vector in $E^6$ with the $\{|i\rangle\}$ basis, one has the (6x6) projection matrices:

$$\Pi^{//}_{i,j} = \sum_{\alpha} \langle i|\alpha\rangle\langle \alpha | j\rangle, \qquad \Pi^{\perp}_{i,j} = \sum_{\beta} \langle i|\beta\rangle\langle \beta | j\rangle,$$

that are explicitly given in the Appendix.

**II.2 Unit cell in $E^6$**

The simplest real structure with icosahedral symmetry is generated by one unique AS (in $E_\perp$) attached to all nodes of $\Lambda$ ($P(1)$ lattice). Its shape is a polyhedron (rhombic triacontahedron) defined by the convex envelope of the projection in $E_\perp$ of the 6d-unit cell. This structure is the canonical icosahedral tiling, very useful for describing the geometrical properties of *i*-QCs (orientational symmetry, quasiperiodic sequences of rows, inflated pentagons). Beyond the understanding of the basic geometrical features of images, they are however of little help in the interpretation of atomic configurations (as given by STM images) that are known to be generated by a complex unit cell composed of several ASs and a non trivial distribution of chemical species within.

Attached to the $F(2)$-lattice, the unit cell in $E^6$ of *i*-AlPdMn, consistent with diffraction data, was found to be decorated by three ASs given in Figure 1: one large triacontahedron $\tau$ times larger than the canonical triacontahedron located at the origin (even nodes $n$ of $\Lambda$[31]), one other truncated triacontahedron of the same size at $(1,0,0,0,0,0)$ (odd nodes $n'$ of $\Lambda$) and a smaller one at $1/2(-1,1,1,1,1,-1)$ ($bc$ nodes: i.e. odd nodes of a $F(2)$-lattice which origin is placed at the body center of the unit cell). Each AS generates an atomic density proportional to its size as given in Table I (see also the Appendix).

With respect to the atomic species distribution, a simple solution acceptable for diffraction data and magnetic properties (giving restrictions on the Mn distribution[32]) is the following:

- Mn distribute on an assembly of small triacontahedra within $n'$-AS.
- Pd distribute on an assembly of small triacontahedra within $n'$-AS and fill the $bc$-AS.
- Al fill the $n$-AS and the remaining volume of $n'$-AS.

Although not fully optimized with respect to diffraction data, because of a small excess in Pd and no relaxation of atomic positions, the above prototypic structure has the great advantage of simplicity and reflects the main features of the atomic order for the icosahedral phase[33].

This structure generates in real space the following Bergman (B) and MacKay (M and M') like clusters with (surprisingly for a ternary compound) a very limited number of chemical configurations. The B-clusters are made of:

- a central Pd atom,
- 12 Al at the vertices of a full icosahedron of radius 0.276 nm
- a dodecahedron of radius 0.407 nm which vertices are decorated by a mixture of the 3 atomic species with an average composition over the 20 sites of 4.96 Mn, 9.71 Pd and 5.33 Al atoms. 9 configurations are only obtained.

M-clusters have one unique chemical configuration:

- 1 Al at center,

- Al on 7 out of 20 vertices of an inner dodecahedron

- 12 Mn at the vertices of a full large icosahedron

- 30 Al at the vertices of a full large icosidodecahedron

M'-clusters belong to 4 chemical configurations:

- M'1 is made of 1 Al atom at the center[33], Al on 7 out of 20 vertices of an inner dodecahedron, 12 Al on a full large icosahedron and 30 Pd on a full large icosidodecaheron and appears with a 47.2 % probability.

Other M' clusters differ only by their outer shell and appear with the following probabilities:

- M'2: 29 Pd and 1 Al on the large icosidodecaheron (23.6 %).

- M'3: 28 Pd and 2 Al on the large icosidodecaheron (11.2 %).

- M'4: 15 Pd, 10 Al and 5 Mn on the large icosidodecaheron (18 %).

On the average 26.8 Pd are present on the large icosidodecahedron giving an average concentration of 21.72% Pd in the bulk.

The decorated unit cell being known, typical real configurations illustrating the resulting structure can be generated by the cut method. Moreover, investigations of the perpendicular space allow extensive descriptions of local generic properties (chemical order, cluster description, quasicristallographic plane densities or step heights). In the following, we propose to examine terrace and step properties as given by the polyhedral model, to be compared with STM observations. Focusing our discussion on QCs with icosahedral symmetry, $5f$, $3f$ and $2f$ surface orientations are considered.

## III. Step and terrace stacking sequences

Like for usual crystals, atomic rows and planes in QCs can be defined as rows and planes which indexes belong to $Z(\tau)$. Each row contains an infinite number of atoms and define a quasi-crystallographic direction. Planes orthogonal to one QC direction and containing atoms define the QC planes. We designate by $5f$, $3f$ and $2f$ planes, the planes perpendicular to the respective high symmetry axis. In the 6d space, quasicrystallographic directions are the trace in $E_{//}$ of a 2-d sublattice of the main 6d lattice $\Lambda$. Similarly, quasicrystallographic planes are the trace in $E_{//}$ of a 4-d sublattice of $\Lambda$.

As for surfaces close to a high symmetry orientation, we note $\hat{z}$ the direction orthogonal to the main terraces. Interplane distances in the bulk along $\hat{z}$ are possible step heights separating adjacent terraces. For the purpose of the analysis of STM images, all possible sequences of elementary step heights are required together with the in plane structure. All necessary information about step height sequences is contained in 2-d planes $E_z^2$ of $E^6$ that are obtained as follows: giving the high symmetry QC direction ($\hat{z}$ axis in $E_{//}$), one defines the associated perpendicular axis $\hat{z}_\perp$ (in $E_\perp$) (see the Appendix). $\hat{z}$ and $\hat{z}_\perp$ together define $E_z^2$ that is further filled by the projection of all ASs (projection of $\Lambda$, with each node decorated by the projection on $\hat{z}_\perp$ of the convex envelope of the associated ASs). In that way $E_z^2$ contains a lattice $\Delta_z$ that is the 2-d projection of the lattice $\Lambda$. We note $\{|z_{//}\rangle, |z_\perp\rangle\}$ the orthogonal basis of $E_z^2 = (\hat{z} \times \hat{z}_\perp)$. We use the generating basis $\{|z_1\rangle, |z_\tau\rangle\}$ of the $n$-node projection as a generating basis of $\Delta_z$. Complementary, one associated QC plane $\mathrm{P}(z, z_\perp)$ at altitude $z$ (along $\hat{z}$ ) is the trace in $E_{//}$ of a 4-d orthogonal subspace in $E^6$. This 4-d subspace decomposes into parallel and perpendicular planes: $\mathrm{P}(z, z_\perp) = \mathrm{P}_{//}(z) + \mathrm{P}_\perp(z_\perp)$.

For determining the basis $\{|z_1\rangle, |z_\tau\rangle\}$ of $\Delta_z(nf)$ for each high symmetry ($5f$, $3f$ or $2f$ ) plane orientations, one use the property that $E_z^2(nf)$ is the invariant plane of $E^6$ upon the selected rotation. From the stereographic projections of the principal symmetry directions[1], the matrix of the symmetry rotation operator can be written. The 2 eigenvectors of this matrix with eigenvalue 1 define one basis of $E_z^2(nf)$. Knowing one first invariant vector $|z0_1\rangle$ of $E^6$, its normalized projections into $E_{//}$ and $E_\perp$ (i.e. along the selected $E_{//}$ and $E_\perp$ symmetry axes) gives one $\{|z_{//}\rangle, |z_\perp\rangle\}$ basis of $E_z^2$. In order to built a generating basis of $\Delta_z(nf)$, it is astute to get the second necessary basis vector of $E_z^2$ by means of the $\tau$ inflation method: $|z0_\tau\rangle = (\tau|z0_{1//}\rangle, (1-\tau)|z0_{1\perp}\rangle)$. The $\tau$ inflation matrix is given in Appendix 1. Because of the $\tau$ inflation, $|z0_1\rangle$ and $|z0_\tau\rangle$ are not collinear, and define by projection on $\hat{z}$ and $\hat{z}_\perp$ a $Z(\tau)$-module on both axes. $|z0_1\rangle$ and $|z0_\tau\rangle$ belong to $\Lambda$, thus forming a good basis of $\Delta_z(nf)$. By linear combinations of $|z0_1\rangle$ and $|z0_\tau\rangle$ the generating $\{|z_1\rangle, |z_\tau\rangle\}$ basis of $\Delta_z$ can be further obtained. We choose the generating basis in such a way that projection of all even nodes of $\Lambda$ $|V\rangle = |n_1, n_2, n_3, n_4, n_5, n_6\rangle$, ( $\sum_i n_i = even$ ) can be written under the form $nz_1|z_1\rangle + nz_\tau|z_\tau\rangle$, where $(nz_1, nz_\tau)$ are integers.

The stacking sequence of atomic planes for each high symmetry orientation can be now easily generated by the standard cut algorithm: at one level $z$ (in $E_\perp$ ). Atoms are present within $\mathrm{P}_{//}(z)$ as soon as this real plane cut the ASs. Therefore, along $\hat{z}$ one QC atomic plane is present each time the cut axis $\hat{z}$ intersects the projection on $\hat{z}_{\perp}$ of the convex envelope of the ASs. For *i*-AlPdMn, the $\Delta_z$ decorated lattice is given Figure 2 for the $5f$ orientations and in Figure 3 and Figure 4 for the $3f$ and $2f$ orientations. Table II gives for the 3 high symmetry orientations, the vector basis $\{|z_1\rangle,|z_\tau\rangle\}$ of the projected $F(2)$ lattice $\Delta_z$ and the chosen step height unit (see below). The basis of the complementary space $\mathrm{P}(z,z_{\perp})$ is further obtained by standard orthogonalization techniques: Explicit details on the determination of the basis of $\Delta_z$ and $\mathrm{P}(z,z_{\perp})$ are given in the Appendix.

Figure 5 illustrates the correspondence between $E_{//}$ and $E_{\perp}$ atom coordinates for the $5f$ orientation. In this figure the grey level of each plane is chosen in relation to the level of cut of the $n$-AS. Generated by the central part of $n$-AS (b) divided in 3 slices, atoms are distributed in real space in well defined $5f$ planes (a). Their associated $\delta z_{\perp}$ coordinates gives points also distributed in well-defined planes (c) that right belong to the corresponding slices.

The $\Delta_z$ representation is useful to extract generic properties of step heights. By a cut of $E_z^2$ by the $\hat{z}$ axis (along $E_{//}$) at various heights (in $E_{\perp}$), one explore all possibilities for step height configurations. Several points are to be noticed:

- In the present polyhedral model, ASs verify the closing condition: all vertices have their coordinates on $Z(\tau)$ that is also valid for edges of projected ASs. This makes also that each edge of a projected AS is aligned with at least one other along the parallel direction.

- $n$, $n'$ and $bc$ nodes project in $\Delta_z$ at different points for the $5f$ and $3f$ orientations (see Figure 2 and Figure 3) whereas they project on top of each other in the $2f$ case (Figure 4): odd order z-directions generate separate $n$ and $n'$ 2d projected lattices and $2f$ directions generate one unique 2d-lattice. Hence, each $5f$ and $3f$ planes is generated by ***one unique*** kind of ASs whereas $2f$ planes contain atoms issued from ($n$, $n'$) or ($n$, $n'$, $bc$) ASs.

- The frequency of each family of plane (issued from $n$, $n'$ and $bc$ ASs) is related to the height $h_{AS}$ of the projected ASs (along $\hat{z}_\perp$): more points are generated by the $\hat{z}$ cut from a large ASs than from small ones (see Table III and Appendix).

- Orientational symmetry makes that $|z_1\rangle, |z_\tau\rangle$ are chosen with the length of their $E_{//}$ components related by $\tau$. For each $nf$ orientation, a step height unit $\delta_n$ can be defined (given in Table II). To bear in mind the basis $\{|z_1\rangle, |z_\tau\rangle\}$, it is taken equal to the $E_{//}$ projection of $|z_1\rangle$ i.e. between projected nodes of $\Lambda$ of the same parity. Step heights given by these equivalent ASs are to be found on the $Z(\tau)$-module. Distances along $\hat{z}$ between nearest neighbor ASs define elementary interplane distances that are possible step heights (see Table IV).

For the $5f$ orientation (see Figure 2), and according to the polyhedral structure, one edge of projected $n$-AS is aligned in $E_\perp$ with the other extremity of the $n+(3,-1)$-AS (expressed in the $\{|z_1\rangle, |z_\tau\rangle\}$ basis). Upper edge of $n+(1/2,0)$ projected $bc$-ASs and $n+(0,1/2)$ $n'$ ASs are both aligned with the lower edge of $n+(3,0)$ projected $n$-AS. The interplane distances between $n$ and $bc$-planes (issued from AS attached to respectively $n$ and $bc$ nodes) are on the $Z(\tau)+1/2$ module (i.e. are of the form: $((2n+1)/2+m\tau)\,\delta_5$) The interplane distances between $n$- and $n'$-planes are on the $Z(\tau)+\tau/2$ module (i.e. are of the form: $(n+(2m+1)\tau/2)\,\delta_5$). and between $bc$-$n'$ planes on the $Z(\tau)/2$ module (i.e. are of the form: $((2n+1)+(2m+1)\tau)\,\delta_5/2$). The distance between adjacent $n$-AS planes are on the $Z(\tau)$ module. Observation of step heights with half irrational or half integer indices indicates that the upper and lower terraces belong to different ASs and are not isomorphic. According to the AS closing rule, 2 or 3 interplane distances only are generated from each AS (see Table IV). This makes 1 to 4 possible elementary step heights between adjacent ASs. Along $\hat{z}$, the occurrence probability of each elementary distance is given by the common facing height in $E_\perp$ of corresponding ASs. Heights in $\delta_n$ units and probabilities are given in Table IV, and the histogram of distances in Figure 6.

For $3f$ orientation, 5 elementary interplane distances are generated (see Table IV). The distance between $n$-$n'$ planes are on the $Z(\tau)/2$ module (i.e. are of the form: $((2n+1)+(2m+1)\tau)\,\delta_3/2$) and between $n$-$bc$ planes on the $Z(\tau)+\tau/2$ module. Step heights with a half irrational index indicates that the adjacent terraces are non isomorphic.

For $2f$ orientation, $n$, $n'$ and $bc$ projected lattices are superimposed. Projected $n$-As and $n'$-AS are of the same size. Atoms issued from the three ASs are present for a cut close to the nodes whereas $n$ and $n'$-ASs only contribute for cuts at a distance greater than $\sqrt{(2+\tau)/10}$ to the node. All interplane distances belong to the $Z(\tau)$ module.

## IV. In plane terraces

The 2-d decorated lattices $\Delta_z$ allow a description of the stacking plane properties. Once a plane is chosen at altitude z where one projected AS is cut at a distance $\delta z_\perp$ to the node (with $(z, z_\perp)$ coordinates in $E_z^2$), the in plane structure is obtained by the cut algorithm within $\mathrm{P}(z)$ along $\mathrm{P}_{//}(z)$. The acceptance windows (along $\mathrm{P}_\perp(z_\perp)$) for each type of nodes are **the sections** in $E_\perp$ (polygons) of the 3d-ASs at the given altitude $\delta z_\perp$. The in plane density is proportional to the area $S_{AS}(\delta z_\perp)$ of the AS sections. ASs being regular polyhedron, the AS can be decomposed in successive facetted slices for each symmetry orientation. In-plane atomic density $\rho S_{AS}(\delta z_\perp)$ decreases quadratically with $\delta z_\perp$ within each slice (the normalization method is given in the Appendix).

Along the $5f$ orientation, 7 slices of the outer rhombic triacontahedron ($n$ and $bc$-ASs) can be defined (See Figure 7). It is noticeable that the widest central part is a decagonal right-angled prism giving the constant maximal density. All planes $(z, z_\perp)$ where the cut fall within this slice gives locally isomorphic planes with local 10-fold patterns. The area of the central slice for $n$-AS is $\tau^4$ times larger than for $bc$-AS and gives the maximally dense planes. We call this set of maximally dense equivalent $n$-planes the primary terraces of the structure. On each side of the central slice, the density decays quadratically across the next slices, to eventually being very small for the outermost slices (see Figure 8). Typical configurations of terraces resulting from a cut of the $n$-AS within the successive slices are given Figure 9. For $n'$-AS (truncated triacontahedron) density variations close to the $n$-AS's one are obtained.

Similarly, a rhombic triacontahedron can be decomposed in 9 slices along the $3f$ orientation, with a central slice of constant surface and in 6 slices for the $2f$ orientation. Figure 10 gives the density variation in those cases.

From surface configurations generated by the 6-d polyhedral model, electronic density maps can be generated from the simple superposition of exponentially decaying electronic densities centered on atoms $e^{-\kappa d}$. In our calculation, the chosen decay constant $\kappa = 10$ nm$^{-1}$ is assumed to be the same for all atoms. All planes below the terrace plane contributing to the electron density at the view distance ($d_z$ = 0.35 nm) are included. View distance and decay constant are adjusted so that smooth images where atoms are still visible are obtained. Figure 11 shows such images for $5f$-planes issued from each slice of the $n$-AS. Images of other planes (at different $z$) obtained from cuts at the very same relative heights $\delta z_{\perp}$ within each slice are given in Figure 12. For the two set of images, the density of surface atoms is the same and their differences result from the interplane distances and density of the underneath planes. More open surfaces (presumably of higher energy) are obtained from ASs with distant (in $E_{//}$) or low density underneath AS at the cut level. It was shown that local configurations of high-resolution STM images can be identified to such configurations issued from the polyhedral QC model[9].

## V. Experimental: He diffraction

With the aim of comparison of structural models with surface structures the 5-fold surface of i-AlPdMn[9] (exact composition: $Al_{71.1}Pd_{20.2}Mn_{8.7}$, as measured by plasma optical emission spectroscopy at different locations) has been studied. Prior to STM study, giving direct views of local configurations to be compared with QC models, surface preparation was followed by LEED, Auger Spectroscopy and He diffraction as a highly sensitive tool to surface morphology. Experimentally, as for complex materials, the surface cleaning process must be investigated with care to get the best equilibrium surface state (chemical composition and crystallinity). He diffraction (energy of incident neutral He atoms: 21 meV, wave vector: $K_i = 65\ nm^{-1}$) is a technique only sensitive to the very outermost surface plane that allows to characterize the morphology of the top most atomic plane on the atomic scale over a large sample area. Its high sensitivity to surface defects and the ability to monitor the surface during temperature variation, atom deposition or $Ar^+$ bombardment are other advantages of the technique. Up to now, He diffraction was observed on the fivefold surface of i-AlPdMn[9 34 35] and on the 2-fold and 10-fold surfaces of d-Al-Ni-Co[34 35 36]. In the following some details on this preliminary experiment are given.

Under UHV, one as-grown facet of the sample has been softly mechanically polished in order to remove the native oxide layer. There, a standard cleaning process (cycles of 5 µA, 400 V, 1 h $Ar^+$ sputtering prior to 1h, 925 K annealing) was used and monitored by Auger electron spectroscopy (AES) and low energy electron diffraction (LEED) (see Figure 13). For the well annealed surface, the intensity ratio of the AES peaks are Al (68 eV)/Pd (330 eV) = 1 and Mn (40 eV)/Pd(330 eV) = 0.3. The LEED pattern presents sharp spots with fivefold symmetry. Upon short $Ar^+$ sputtering, the relative intensity of the Al peak decreases showing a preferential Al sputtering in favor of an Al rich surface composition.

During the 3 months of our experiment (more than 100 $Ar^+$ bombardment-annealing cycles), the very same He diffraction pattern is obtained with a slightly increasing He reflectivity showing the stability of the surface upon cumulated surface preparation cycles. The He diffraction pattern exhibits a narrow and intense specular peak. An excellent reflectivity, up to 6% of the incident beam can be reached (see Figure 14). This intense specular peak is surrounded by a fivefold set of diffraction peaks of low intensity ($I_G \cong 1.5$ % of the specular intensity). In the incidence plane, position of successive diffraction peaks follow a $\tau$ scale: $\Delta K_G = \tau^{G-1} \Delta K_1$, where $\Delta K_1 = 2\pi / a$, a = 1.7 ± 0.1 nm.

Versus time, a slow attenuation of the He reflectivity is observed at room temperature. AES allows detecting oxygen (510 eV peak) contamination after few hours showing the high reactivity of these Al based samples that imposes high-quality UHV conditions ($\leq 10^{-10}$ Torr). During long experiments, measurements are performed at $T$ = 473 K where $I_0$ can be kept constant over several hours.

Specular and diffraction peak intensities show slow variations versus incident angle and a constant specular width (See Figure 15). This indicates that on the average the surface is likely made of flat terraces without too many steps. Taking into account the instrumental resolution, the residual specular width is constant and equal to 0.013 ($2\pi$/a), giving a surface coherence length higher than 50 nm that will be confirmed by the STM experiment. The $I_G / I_0$ ratio being small, an easy estimate of the terrace corrugation can be obtained using the kinematic (or eikonal) approximation for the He diffraction intensities. The amplitude $Ac$ of the fivefold Fourier component of the surface corrugation is given by:

$$A_c = \frac{1}{K_{G\perp} - K_{0\perp}} \sqrt{\frac{I_G \; K_{0\perp}}{I_0 \; K_{G\perp}}}$$

where $K_{0\perp}$ (specular) and $K_{G\perp}$ (diffraction peaks) are the perpendicular to the surface components of the wave vector. One obtains $A_c$ = 1.92 pm: terraces are flat dense planes with very low corrugation.

Measurements of the specular intensity ($I_0$) versus temperature allow estimating the thermal attenuation (Debye waller) factor. Due to thermal vibration of surface atoms, the He diffraction peak intensities decay smoothly with temperature $T$ (see Figure 16). As for stable dense metallic surfaces, the thermal attenuation (Debye Waller factor) is:

$$F(T) = e^{-\Delta K_{\perp}^2 \left\langle u_{\perp}^2 \right\rangle_T}$$

where $\Delta K_{\perp} = K_{f\perp} - K_{i\perp}$ is the perpendicular momentum transfer including refraction within the He-surface potential well (well depth $D$):

$$K_{i,f\perp} = \sqrt{\left[K \cos\left(\theta_{i,f}\right)\right]^2 + \frac{2M_{He}D}{\hbar^2}},$$

where $M_{He}$ is the mass of an He atom and $\hbar$ the Boltzmann's constant. $\left\langle u_{\perp}^2 \right\rangle_T$ is the He-surface potential mean square displacement normal to the surface (that must be slightly lower than the surface atom vibration amplitude). According to the Debye approximation $\left\langle u_{\perp}^2 \right\rangle_T = A \cdot T$, $F(T)$ is exponential for one given scattering condition. Anharmonicity could make the thermal attenuation higher at high $T$. For low T, the measured exponential $I_0$ decay gives $A$ = 2.1 $10^{-7}$ $nm^2/K$, comparable to values measured for dense faces like Cu(001). The Debye Waller factor for Cu, including the high temperature anharmonicity is reported in Figure 16 for comparison. Clearly, $I_0$ deviates significantly in a reversible way from normal thermal attenuation above 750 K. On the contrary, the relative intensity of the diffraction peaks increases (see Figure 17). Anharmonicity would affect similarly adjacent diffraction peaks and one may conclude that the surface morphology is reversibly altered above 750 K Formation of surface point defects or phasons with $E_{//}$ component may account for these observations.

On Figure 16, the evolution of the specular intensity during annealing after $Ar^+$ bombardment is also reported. It appears that the surface does not order below 650 K and is fully restored after annealing above 825 K. This range of temperature is in agreement with other observations[37] and coincides with the temperature where the anomalous thermal attenuation is observed. Restoring the surface structure needs restoring its composition that requires an active bulk diffusion. It is noticeable that surface ordering occurs above the transition between two bulk diffusion regimes observed for Pd in i-AlPdMn[38]. Below 725 K, diffusion occurs by a mechanism assisted by collective phason flips, whereas vacancy mediated bulk-diffusion is pointed out above.

He diffraction was the perfect tool to investigate the surface preparation of our sample. Wide dense 5-fold flat terraces on an homogeneous surface are observed where a direct view of some configurations are safely expected to be seen in the next STM study.

## VI. Experimental: STM study

Atomic resolution with STM is not so easy to obtain on metallic sample where the surface corrugation of dense planes is generally close to the resolution limit. Up to now high quality STM images of QCs with atomic resolution has been obtained on *i*-AlPdMn[9 24 25] and *i*-AlCuFe[14] that belongs to the same symmetry class. Surprisingly local configurations on high resolution images of these two alloys are clearly different. Interpretation of such images by means of QC models allows interpretation of these differences[39].

The very same cleaning process than for He scattering was used before STM observation of the $5f$ *i*-AlPdMn surface. Like the one presented in Figure 18, ≈100 nm size terraces are present over the sample. Apart some scarce black holes and random white dots (adatoms) terraces are the wide flat and dense terraces expected from the He experiment. The Fourier transform over selected part of the image gives the very same $5f$-pattern than the entire image.

**VI.1. Terrace structure**

In the following, an area of the sample with surface orientation vicinal to the $5f$ axis is explored at various resolutions. Such a vicinal surface (See Figure 19) allows exploring step height sequences and the structure of one terrace together with those of adjacent terraces that are the emergence of upper and lower $5f$ planes. Various heights on the $Z(\tau)$-module separate adjacent terraces (see Figure 20). Images at higher resolution (Figure 21) within the same area reveal the terrace atomic surface structure. Within one terrace, all hollow (or filled) flowers have the same orientation. As shown in Figure 21, same flowers with the very same orientations are also observed on adjacent wide terraces. This correlation shows that long range QC order is well present in the entire image and indicates that wide terraces belongs in some way to isomorphic planes of the QC structure.

Zooming within one wide flat terrace (see Figure 21 and Figure 22) allowed reaching the STM ultimate resolution. The atomic pattern is mainly composed of elementary 'donuts' that are organized to form filled or empty $5f$-flowers. Similarly to Figure 18, the Fourier transforms (Figure 23) of area within the very same wide central terrace of images of Figure 19, Figure 21 and Figure 22 exhibits well defined 10-fold patterns. Intense peaks are present within the range $\pi / 0.6$ nm$^{-1}$- $\pi / 0.14$ nm$^{-1}$ that is the typical donuts and flowers size. Above this scale, no long range corrugation of terraces is pointed out.

Identification of the terrace structure by means of the 6-d model is a priori a not so easy task. In agreement with AES results, dynamical LEED analyses[40 41 42] of the $5f$-*i*-AlPdMn surface indicate that terraces are Al rich planes. Then, $n$-AS have to be first and foremost considered[9] to interpret images at the highest resolution. Position of atoms within $\mathrm{P}_{//}(z)$ defined by a cut of the $n$-AS at $\delta z_{\perp} = -\tau\,\delta_5$ in $\Delta_z(5f)$ are superimposed to the image of Figure 22. Despite a few supernumerary atoms (in favor of a cut at a slightly higher distance from the node) a fair agreement is obtained in the distribution of donuts and flowers. 3 to 5 atoms contribute to form the donut pattern. The filled and empty flower distribution is also well reproduced. However, the cut level cannot be more accurately fixed in that way. Changing the cut level $\delta z_{\perp}$ around the chosen value, typically inside $[-\tau, -1-\tau]$ (i.e. outside the central decagonal slice and within the 2$^{nd}$-3$^{rd}$ slices) will only slightly change the atom density, the overall pattern remaining for the most part identical (isomorphic terraces).

It is noticeable that the local patterns of the $5f$-*i*-AlPdMn surface obviously differ from those observed at the $5f$-*i*-AlCuFe surface where $10f$ cogwheels with empty or filled center are the most common local configuration within dense terraces[14]. The local average $10f$ symmetry, strongly suggests a possible identification to planes generated by a cut of $n$ or $n'$–AS within their central slices. The different nature of the chemical elements and concentrations between the two alloys would account for the very distinct appearance of their dense terraces, though they belong to the very same icosahedral symmetry.

**VI.2. Steps heights**

As previously shown and discussed in the case of i-AlCuFe[27], height differences between wide terraces on STM images (see for instance Figure 20) are much larger than interplane distances. Indeed, they are much higher than the minimum step height $(-3/2+\tau)\delta_5$ given by the polyhedral model (see Table IV and histograms of Figure 6). $\tau\,\delta_5$, $\tau^2\,\delta_5=(1+\tau)\,\delta_5=0.68$ nm and $\tau^3\delta_5=(1+2\tau)\,\delta_5$ are mainly observed. Thus, height differences between wide terraces result from the composition of several individual steps (bunched steps) that may be separated by very narrow terraces (see Figure 20 and Figure 21). Step bunching antiphase boundaries are also present, where one step leaves a bunch to join another.

It is thus shown that all $5f$ planes do not appear as surface terraces. Interplanar distances must be composed to give the observed heights. For usual fcc and bcc chemically ordered alloys, chemical order leads to step pairing phenomena in such a way that all terraces are low surface energy planes (Cu for Cu-Pd(17%)[12 43] and CuAu for $Cu_3Au$ (step pairing)[12], Al for $Fe_3Al$ (4 steps bunching)[13]). For the cubic structure, vicinal cut of the crystal makes terraces with the same geometry and atomic density but that differs by their chemical composition and consequently by their free surface energy. Dense plane poor in the "segregating species" are of high energy: for the cfc-$L1_2$ $Cu_3Pd$ alloy, Cu is the segregating element and pure Cu terraces are only observed and not Cu-Pd mixed, whereas for $Cu_3Au$, Au segregates and Cu-Au mixed terraces are observed. The total surface free energy is function of the average terrace energy and of the step energy. It was experimentally observed that expanding the width of low energy terraces at the expense of the formation of multiple steps allows minimizing the total surface free energy[12]. In addition, low density planes where numerous bonds are broken giving isolated atoms are usually of high energy and would very unlikely appear as uppermost terraces.

Applying such rules to QCs would allow rejecting low density planes from the stacking sequence and selecting dense planes according to their chemical nature. In that way, widths of high energy terraces shrink and bunching of elementary steps results. Al is found to be the segregating element and terraces must correspond to Al-rich planes in the model (cut of $n$-AS). The maximal projected area in $P(z)$ of the $bc$ ASs being small (see Figure 8), corresponding planes are of low density and can be rejected as possible terraces as well as planes define by a cut in $\Delta_z(5f)$ of the most external slices of the $n$ and $n'$ ASs.

Figure 24 gives a STM image of another vicinal area of the sample together with the height profile along its diagonal (1). The analysis of the image allows distinguishing along the profile the experimental step-main terrace sequence (2). Terrace widths being much smaller than the image size, the sequence cannot be accurately corrected for scanning curvature, this adding some uncertainty on the terrace heights. The height profile is compared to typical profiles issued from the polyhedral model. In the model, the average slope is fixed to that of the STM image and terrace widths are arbitrarily chosen proportional to their in plane atomic density (function of $\delta z_{\perp}$). Several profiles are presented: (3) is the resulting profile when all ASs are considered. (4) is obtained by rejection of the lowest density planes ($bc$-AS and external slices of $n$-AS and $n'$-AS). (5) is obtained if only the dense part of Al-rich $n$-AS are considered (chemical selection). All these profile shapes does not match the experimental one. Dynamical LEED analyses[40 41 42] indicates that a mixed Al-Pd plane of high density is located 0.05 nm below the main terraces that designates (by the distance and chemical composition) $n'$-AS as underlying plane of $n$-AS.

Analysis of recent photodiffraction experiments[44] also conclude in favor of a dense layer lying 0.042 nm below the Al rich surface plane. According to these results, selecting now (6) the dense part of $n$-AS with underneath close plane of high density ($n'$-AS) gives a profile that much better reproduce the experimental one. Note that the absence in the experimental images of wide terraces with $10f$ local patterns (like for $5f$ i-AlCuFe[14]) would strongly suggest rejecting the central decagonal slice of $n$-AS. This criterion removes nearly the same part of the $n$-AS than the condition of having an underneath $5f$-dense plane ($n'$-AS without their external slices).

The step sequence of Figure 20, which central terrace has been attributed to a $n$-AS cut within the 2$^{nd}$-3$^{rd}$ slice, can be compared with sequences of the QC polyhedral model. The bunched step decomposition of Figure 21 must be also included. In a first attempt main terraces are only considered. According to the above arguments, these terraces must belong to $n$-AS in $E_z^2$. Attributing one wide terrace to one $n$-AS, and changing the cut level within the whole extension of the $n$-AS allows exploring all step heights configurations. As discussed above, dense part of $n$-AS is considered with beneath dense $n'$-AS plane. In addition, observation of only $5f$ local patterns within the main terraces (filled and empty flowers) allows slightly reducing the used $n$-AS by rejecting the central decagonal part. This is not mandatory but allows reducing the agreement window for $z_\perp$ to [0.809, 1.191] (see Figure 25) where the experimental sequence is well reproduced. The level of cut is within the same range than expected from the identification of the local configuration within the terrace of Figure 22 $(z, \delta z_\perp) = (1+\tau, -\tau)\,\delta_5$. It is noticeable that main terraces all belong to $n$-AS (Al rich planes) with a cut level relatively to each node satisfying both criteria of dense terraces with underlying dense plane. On the contrary, all narrow intermediate terraces belong to other ASs.

In conclusion to this analysis, the 6d polyhedral model allows to give interpretation of the step stacking sequences observed at QC surfaces. In contrast with ref[27] where step height sequences are interpreted in term of a Fibonacci sequence of thin and thick blocks of atoms, the polyhedral model, where the distribution of chemical species is fixed, allows a deeper understanding of the STM observations. In that way, the plot of $\Lambda_Z$ map appears very useful for a full description of the observed atomic configurations. With this model, it is remarkably found that a step sequence allows defining the level of cut in fair agreement with that obtains from the analysis of an image at the atomic level of one terrace structure. As previously discussed for i-AlPdMn in Ref[45], the present identification of the most common configurations within wide terraces (from both step sequences and high resolution STM images) shows that they do not correspond to the maximally dense planes of the structure (cut of $n$-AS out of the decagonal central slice.

## VII. Conclusion

In conclusion, it is shown that the polyhedral model is an essential tool to investigate quasicrystal surfaces and real images cannot be interpreted without their localization in the 6-d space. Exploration of the perpendicular space is the way to investigate all possible local configurations. Such models provide information on terrace structures with their chemical composition, step heights and terrace-step stacking rules. In the following Appendix, all necessary basis for application to the analysis of surface configurations are given.

STM images of the $5f$ *i*-AlPdMn surface show mainly than only some atomic planes extend as terraces at the surface. Similarly to usual *cfc* and *bc* ordered alloys, the chemical nature of the plane must be considered. In addition for QC, the in plane and sub-plane densities appear as another relevant criteria. Based on neutron and X-ray diffraction data, the present surface investigation allows drawing a steady model for the *i*-QC structure. Further identification of the emergence at the surface of structural defects is another challenge.

# Appendix

## 1. Basis of the 6-d space

Using the indexing notations proposed by Cahn, Shechtman and Gratias (CSG)[1] the coordinates of the orthonormal basis vectors $|i\rangle$ of $E^6$ in $E_{//}$ ($\{|\alpha\rangle\}$ basis) and $E_{\perp}$ $\{|\beta\rangle\}$ basis (i.e. scalar products $\langle\alpha|i\rangle$ and $\langle\beta|i\rangle$) are given in Table A-I.

## 2. Projection matrices in $E_{//}$ and $E_{\perp}$

The projection matrices from $E^6$ ($\{|i\rangle\}$ basis) to $E_{//}$ ($\{|\alpha\rangle\}$) or $E_{\perp}$ ($\{|\beta\rangle\}$) are:

$$M^{//}_{\alpha,i} = K\begin{bmatrix} 1 & \tau & 0 & -1 & \tau & 0 \\ \tau & 0 & 1 & \tau & 0 & -1 \\ 0 & 1 & \tau & 0 & -1 & \tau \end{bmatrix} \text{ and } M^{\perp}_{\beta,i} = K\begin{bmatrix} -\tau & 1 & 0 & \tau & 1 & 0 \\ 1 & 0 & -\tau & 1 & 0 & \tau \\ 0 & -\tau & 1 & 0 & \tau & 1 \end{bmatrix}$$

with $K$ the normalization factor. Resulting coordinates $\langle\alpha|V\rangle = h + h'\tau, k + k'\tau, \ell + l'\tau$ in $E_{//}$ and $\langle\beta|V\rangle = h' - h\tau, k' - k\tau, \ell' - \ell\tau$ in $E_{\perp}$ gives the indexes of the CSG description: $(h/h', k/k', \ell/\ell')$[1].

The coordinates in $E^6$ $\{|i\rangle\}$ of a vector of $E_{//}$ ($\{|\alpha\rangle\}$ basis) (or $E_{\perp}$ ($\{|\beta\rangle\}$)) are obtained with the transposed matrices:

$$C6_{//} = {M^{//}_{\alpha,i}}^{T} \qquad C6_{\perp} = {M^{\perp}_{\beta,i}}^{T}$$

and coordinates in $E^6$ $\{|i\rangle\}$ from CSG indexes are given by:

$$C6_{//}\begin{pmatrix} h+h'\,\tau \\ k+k'\,\tau \\ \ell+\ell'\,\tau \end{pmatrix} + C6_{\perp}\begin{pmatrix} h'-h\,\tau \\ k'-k\,\tau \\ \ell'-\ell\,\tau \end{pmatrix}$$

The projection matrices giving the coordinates in $E^6$ ($|i\rangle$ basis) of the projection of a vector into $E_{//}$ or $E_{\perp}$ are:

$$\Pi^{//} = \frac{2\tau-1}{10}\begin{bmatrix} 2\tau-1 & 1 & 1 & 1 & 1 & -1 \\ 1 & 2\tau-1 & 1 & -1 & 1 & 1 \\ 1 & 1 & 2\tau-1 & 1 & -1 & 1 \\ 1 & -1 & 1 & 2\tau-1 & -1 & -1 \\ 1 & 1 & -1 & -1 & 2\tau-1 & -1 \\ -1 & 1 & 1 & -1 & -1 & 2\tau-1 \end{bmatrix}$$

$$\Pi^{\perp} = 1-\Pi^{//} = \frac{1-2\tau}{10}\begin{bmatrix} 1-2\tau & 1 & 1 & 1 & 1 & -1 \\ 1 & 1-2\tau & 1 & -1 & 1 & 1 \\ 1 & 1 & 1-2\tau & 1 & -1 & 1 \\ 1 & -1 & 1 & 1-2\tau & -1 & -1 \\ 1 & 1 & -1 & -1 & 1-2\tau & -1 \\ -1 & 1 & 1 & -1 & -1 & 1-2\tau \end{bmatrix}$$

## 3. Basis of $\Delta_z$ and $P_z$ associated to each high symmetry QC axis

### *3.1. Outline method*

The hyperplane associated to one direction $E_z^2 = (\hat{z}+\hat{z}_{\perp})$ contains the lattice $\Delta_z$, projection of $\Lambda$ in $E_z^2$. The 4-d perpendicular plane at altitude $z$ is noted: $P(z) = P_{//}(z) + P_{\perp}(z)$. The basis of these subspaces are obtained as following:

One normalized vector $|V_1\rangle$ of $E^6$ along the chosen axis is first chosen. A second vector giving a basis of $E_z^2$ and allowing to generate a $Z(\tau)$ module in both subspaces $E_{//}$ and $E_\perp$ can be obtained by the $\tau$ inflation method: $|V_\tau\rangle = \left(\tau\,\Pi^{//} + (1-\tau)\,\Pi^\perp\right)|V_1\rangle$. The associated matrix is:

$$\tau_{inflat} = \tau\,\Pi^{//} + (1-\tau)\,\Pi^{//} = \frac{1}{2}\begin{bmatrix} 1 & 1 & 1 & 1 & 1 & -1 \\ 1 & 1 & 1 & -1 & 1 & 1 \\ 1 & 1 & 1 & 1 & -1 & 1 \\ 1 & -1 & 1 & 1 & -1 & -1 \\ 1 & 1 & -1 & -1 & 1 & -1 \\ -1 & 1 & 1 & -1 & -1 & 1 \end{bmatrix}$$

and, like $\tau^2 = 1 + \tau$, one has $\tau_{inflat}{}^2 = \mathrm{I}d + \tau_{inflat}$

The generating basis $\{|z_1\rangle, |z_\tau\rangle\}$ of $\Delta_z$ is further built by a linear combination $|z_1\rangle = (a + b\,\tau_{inflat})\,|V\rangle$ of the 2 vectors $|V\rangle$ and $\tau_{inflat}\,|V\rangle$. The basis vectors remains related by the $\tau$ inflation providing: $|z_\tau\rangle = (b + (a+b)\,\tau_{inflat})\,|V\rangle$.

The basis of $\mathrm{P}(z)$ can be built as follows: one first vector $|\pi_1\rangle$ is chosen perpendicular to the 2 basis vectors $\{|z_1\rangle, |z_\tau\rangle\}$. The 3-d vector product of the projection into $E_{//}$ and $E_\perp$ of $|\pi_1\rangle$ and $|z_1\rangle$ gives the components of a second vector $|\pi_2\rangle$ which coordinates in $E^6$ are obtained using $C6_{//}$ and $C6_\perp$ matrices. The last necessary 2 vectors are obtained by $\tau$ inflation of $|\pi_1\rangle$ and $|\pi_2\rangle$. For high symmetry orientations, the 4 independent vectors perpendicular to $\{|z_1\rangle, |z_\tau\rangle\}$ can be easily intuitively built.

### *3.2.* $5f$ *orientation*

The $\{|z_1\rangle, |z_\tau\rangle\}$ basis for the $5f$ orientation is obtained as follows: $|V_1\rangle = (1,0,0,0,0,0)$ is one element of $\Lambda$ along a $5f$ axis[1] that can be chosen as a first vector. Its projections in $E_{//}$ and $E_\perp$ give the orientations of the $5f$ perpendicular and parallel axes giving the orthonormalized basis $\{|z_{//}\rangle, |z_\perp\rangle\}$ of $E_z^2(5f)$. The projection matrix of a $E^6$ vector to the plane define by this basis is:

$$\Pi z_{5f} = \frac{2\tau - 1}{5\sqrt{2}} \begin{bmatrix} 2\tau - 1 & 1 & 1 & 1 & 1 & -1 \\ 2\tau - 1 & -1 & -1 & -1 & -1 & 1 \end{bmatrix}$$

The half height of the projected canonical AS surface is: $\tau/\sqrt{2}$.

By $\tau$ inflation of $|V_1\rangle$, one gets the second vector: $1/2\,(1,1,1,1,1,-1)$. Linear combination of the 2 vectors gives the orthonormalized basis $\{|z0_1\rangle = (1,0,0,0,0,0), |z0_\tau\rangle = 1/\sqrt{5}(0,1,1,1,1,-1)\}$ (these vectors are not projections of $n$-nodes). Projection in $E_z^2(5f)$ of one vector of $\Lambda$ $|V\rangle = |n_1, n_2, n_3, n_4, n_5, n_6\rangle$ is:

$$|V_z\rangle = |z0_1\rangle\langle z0_1|V\rangle + |z0_\tau\rangle\langle z0_\tau|V\rangle = n_1|z0_1\rangle + \frac{n_2 + n_3 + n_4 + n_5 - n_6}{\sqrt{5}}|z0_\tau\rangle$$

One would further write one basis that generates the sublattice of $\Delta_z(5f)$ that is the projection of $n$-nodes (which coordinates in $E^6$ satisfy the condition $\sum_{i=1,\,6} n_i$ is even). One good basis is:

$$\left( |z_1\rangle = |z0_1\rangle - \frac{1}{\sqrt{5}}|z0_\tau\rangle = \frac{1}{5}(5,-1,-1,-1,-1,1) \ , \ |z_\tau\rangle = \frac{2}{\sqrt{5}}|z0_\tau\rangle = \frac{2}{5}(0,1,1,1,1,-1) \right)$$

which vectors are well related by the $\tau$ inflation. With this basis, the projection matrix in $\Delta_z(5f)$ is:

$$\Pi' z_{5f} = \begin{bmatrix} 1 & 0 & 0 & 0 & 0 & 0 \\ 1/2 & 1/2 & 1/2 & 1/2 & 1/2 & -1/2 \end{bmatrix}$$

Translation in $E^6$ of the $n$-node lattice by $1/2(1,1,1,1,1,1)$ gives the $bc$ lattice and by $(0,1,0,0,0,0)$ the $n'$-node lattice. The following relations:

$$\Pi' z_{5f}\left[\frac{1}{2}(1,1,1,1,1,1)\right] = \begin{pmatrix} 1/2 \\ 1 \end{pmatrix} \text{ and } \Pi' z_{5f}(0,1,0,0,0,0) = \begin{pmatrix} 0 \\ 1/2 \end{pmatrix},$$

show that projected nodes are distinct. The lattice of projected $bc$-nodes is obtained by a translation of $1/2|z_1\rangle$ of the $n$ lattice and of $1/2|z_\tau\rangle$ for the $n'$ lattice. (Note that an orthogonal basis in $\{z_{//}, z_\perp\}$of the $n'$ lattice can be also used instead of the chosen basis that is not orthogonal).

Each node of the full lattice in $\Delta_z(5f)$ must be further decorated by the projection of the convex envelope of the ASs. The half height of projected $n$-AS is given in Table II. $n'$-AS being truncated along the 5-fold direction, their projected size is slightly smaller than for $n$-AS.

One trivial solution for a 4-d basis of $5f$ planes is:

$$\{\,(0,-1,1,0,0,0);\,(0,0,0,1,-1,0);\,(0,1,1,-1,-1,0);\,(0,1,1,1,1,4)\,\}$$

### *3.3.* $3f$ *orientation*

$|z0_1\rangle = 1/\sqrt{6}(1,1,1,1,1,1)$ is along one $3f$ symmetry axis of $E^6$. Its projection in $E_{//}$ and $E_\perp$ gives the following orthonormal basis $\{z_{//}, z_\perp\}$ of $E_z^2(3f)$:

$$\sqrt{\frac{7-4\tau}{30}}\{(1+2\tau \quad 1+2\tau \quad 1+2\tau \quad 1 \quad 1 \quad 1), (1 \quad 1 \quad 1 \quad -(1+2\tau) \quad -(1+2\tau) \quad -(1+2\tau))\}$$

which coordinates are the lines of the projection matrix. The half height of the projected canonical AS surface is: $\sqrt{\frac{3(2+\tau)}{10}}$.

After $\tau$ inflation, one gets the second normalized vector $1/\sqrt{3}(1,1,1,0,0,0)$ and by linear combination of the two vectors the orthonormal basis $\{|z0_1\rangle = 1/\sqrt{6}(1,1,1,-1,-1,-1), |z0_\tau\rangle = 1/\sqrt{6}(1,1,1,1,1,1)\}$. With respect to this basis, projection of $|V\rangle$ of $E^6$ in $\Delta_z$ is:

$$|V_z\rangle = \frac{n_1+n_2+n_3-n_4-n_5-n_6}{\sqrt{6}}|z0_1\rangle + \frac{n_1+n_2+n_3+n_4+n_5+n_6}{\sqrt{6}}|z0_\tau\rangle$$

Sum of $n_i$ coordinates for $n$-nodes being even, one gets one orthogonal basis generating the sublattice (of $\Delta_z(3f)$) of the projection of $n$-nodes:

$$\left( |z_1\rangle = \sqrt{\frac{2}{3}}|z0_1\rangle = \frac{1}{3}(1,1,1,-1,-1,-1) \ , \ |z_\tau\rangle = \sqrt{\frac{2}{3}}|z0_\tau\rangle = \frac{1}{3}(1,1,1,1,1,1) \right)$$

which vectors are related by the $\tau$ inflation. The associated projection matrix from $E^6$ is:

$$\Pi' z_{3f} = \frac{1}{2}\begin{pmatrix} 1 & 1 & 1 & -1 & -1 & -1 \\ 1 & 1 & 1 & 1 & 1 & 1 \end{pmatrix}$$

With $\Pi' z_{3f}\left[\frac{1}{2}(1,1,1,1,1,1)\right]=\begin{pmatrix}0\\3/2\end{pmatrix}$ and $\Pi' z_{3f}(0,1,0,0,0,0)=\begin{pmatrix}1/2\\1/2\end{pmatrix}$, projection of $n'$ and $bc$ nodes are distinct from $n$-nodes. The heights of projected $n$-AS and $n'$-AS are the same: $\sqrt{\frac{7+11\tau}{10}}$ and projected $bc$-As are $\tau^2$ times smaller.

One trivial solution for a 4-d basis of $3f$ planes is:

$$\{\ (1,-1,0,0,0,0);\ (0,0,0,1,-1,0);\ (1,1,-2,0,0,0);\ (0,0,0,1,1,-2)\ \}$$

### *3.4.* $2f$ *orientation*

$|z0_1\rangle = 1/\sqrt{2}(1,1,0,0,0,0)$ is along one $2f$ symmetry axis. Its projection in $E_{//}$ and $E_\perp$ gives the following orthonormal basis $\{z_{//}, z_\perp\}$.:

$$\sqrt{\frac{3-\tau}{10}}\{(\tau \quad \tau \quad 1 \quad 0 \quad 1 \quad 0),(1 \quad 1 \quad -\tau \quad 0 \quad -\tau \quad 0)\}$$

which coordinates are the lines of the projection matrix. The half height of the projected canonical AS surface is: $\sqrt{\frac{3+4\tau}{10}}$ .

After $\tau$ inflation, one gets the second normalized vector $1/2(1,1,1,0,1,0)$ and the orthonormal basis $\{|z0_1\rangle = 1/\sqrt{2}(1,1,0,0,0,0), |z0_\tau\rangle = 1/\sqrt{2}(0,0,1,0,1,0)\}$. With respect to this basis, projection of $|V\rangle$ of $E^6$ is:

$$|V_z\rangle = \frac{n_1+n_2}{\sqrt{2}}|z0_1\rangle + \frac{n_3+n_5}{\sqrt{2}}|z0_\tau\rangle$$

One orthogonal basis generating the sublattice (of $\Delta_z(2f)$) of the projection of $n$-nodes is:

$$\left( \ |z_1\rangle = \frac{1}{\sqrt{2}}|z0_\tau\rangle = \frac{1}{2}(0,0,1,0,1,0),\ \ |z_\tau\rangle = \frac{1}{\sqrt{2}}|z0_1\rangle = \frac{1}{2}(1,1,0,0,0,0) \right)$$

and the associated projection matrix from $E^6$ is: $\Pi' z_{2f} = \begin{pmatrix} 0 & 0 & 1 & 0 & 1 & 0 \\ 1 & 1 & 0 & 0 & 0 & 0 \end{pmatrix}$.

With $\Pi' z_{2f}\left[\frac{1}{2}(1,1,1,1,1,1)\right] = \begin{pmatrix} 1 \\ 1 \end{pmatrix}$ and $\Pi' z_{2f}(0,1,0,0,0,0) = \begin{pmatrix} 0 \\ 1 \end{pmatrix}$, projected $n$, $bc$ and $n'$ lattices are superimposed. The heights of projected $n$-AS and $n'$-AS are the same: $\sqrt{\frac{7+11\tau}{10}}$ and projected $bc$-AS are $\tau^2$ times smaller: $\sqrt{\frac{2+\tau}{10}}$.

One trivial solution for a 4-d basis of $2f$ planes is:

$$\{\ (1,-1,0,0,0,0);\ (0,0,0,1,01,-1);\ (0,0,1,0,-1,0);\ (0,0,0,1,0,1)\ \}$$

**4. Atomic densities**

According to the cut algorithm, the bulk atomic density is the ratio of the AS volume $V_{AS}$ to the 6-d hypervolume of the unit cell. The volume of a triacontahedron of side 1 and half height $\tau$ is $4\sqrt{3+4\tau}$ giving an atomic density generated by the canonical AS of height $\tau/\sqrt{2}$ on the $P(1)$ lattice: $\rho at_{AS\ canon.} = V_{AS}/1^6 = \sqrt{2(3+4\tau)}$. For the $F(2)$ lattice, subset of $P(1)$ for which $\sum_i n_i$ =even, the number of nodes[46] is divided by 2 and the contribution of each AS (of volume $V_{AS}$ in $E_\perp$) to the total atomic density is: $\rho at_{AS} = V_{AS}/2$ [atoms/$(E^6\ unit)^3$].

Similarly within $\Delta_z$, the density of QC planes generated by each AS along the $z$ direction is: $\rho_{AS}^{z} = h_{AS} / S_{\Delta z}$ [plane/ $E^6$ $unit$ ] where $S_{\Delta z}$ is the surface of the unit cell $(|z_1\rangle, |z_\tau\rangle)$ that generate the sublattice of the projection of $n$-nodes.

The in plane density is obtained as following: the contribution of each AS to the total bulk atomic density is the product of the density of plane along the chosen QC direction $\rho_{AS}^{z}$ by the average in plane density $\langle \rho S_{AS} \rangle$.

$$\rho_{AS}^{z} \langle \rho S_{AS} \rangle = \rho at_{AS} = V_{AS} / 2 \quad [\text{atom}/ (E^6\ unit)^3 ]$$

One gets: $\langle \rho S_{AS} \rangle = \dfrac{S_{\Delta z}}{2h_{AS}} V_{AS}$ and the AS being uniformly cut over their full height, the atomic density $\rho S_{AS}(\delta z_\perp)$ of a QC plane at altitude $z$, issued from the cut of an AS at a distance $\delta z_\perp$ to the node is:

$$\rho S_{AS}(\delta z_\perp) = \frac{S_{\Delta z}}{2h_{AS}} V_{AS} \frac{S_{AS}(\delta z_\perp)}{\langle S_{AS} \rangle}$$

with the average AS surface: $\langle S_{AS} \rangle = V_{AS} / h_{AS}$, one obtains:

$$\rho S_{AS}(\delta z_\perp) = \frac{S_{\Delta z}}{2} S_{AS}(\delta z_\perp) \quad [\text{atoms}/ (E^6\ unit)^2 ].$$

| | Volume in $E_\perp$ $\left(E^6\ unit\right)^3$ | Volume in $E_\perp$ *bc*-AS unit $=\sqrt{2(7-4\tau)}\ \left(E^6\ unit\right)^3$ | Atomic density $at/\left(E^6\ unit\right)^3$ (see Appendix) |
|---|---|---|---|
| Cubic P-lattice Canonical –AS | $\sqrt{2(3+4\tau)}$ | $\tau^3$ | $\sqrt{2(3+4\tau)}$ |
| $F(2)$ lattice *bc*-AS | $\sqrt{2(7-4\tau)}$ | 1 | $\sqrt{(7-4\tau)/2}$ |
| *n*-AS | $\sqrt{2(47+76\tau)}$ | $\tau^6$ | $\sqrt{(47+76\tau)/2}$ |
| *n'*-AS | $\sqrt{2(43+44\tau)}$ | $5+6\tau$ | $\sqrt{(43+44\tau)/2}$ |
| Total ASs volume | $\sqrt{2(203+244\tau)}$ | $11+14\tau$ | $\sqrt{(203+244\tau)/2}$ |

**Table I: Volume of the different AS and resulting atomic densities (see Appendix). With** $a$ **= 0.6451 nm [**$E^6\ unit$**] for i-**$Al_{71.1}Pd_{20.2}Mn_{8.7}$ **the atomic density given by the model is 4.861 g/cm**$^3$**. The (arbitrary) scale in** $E_\perp$ **can be chosen so as the volume of bc-AS = 1. In that case, the** $E_\perp$ **and** $E_{//}$ **scales are different.**

| | Unit cell of $\Delta_z$ (coordinates in the $\{\|i\rangle\}$ basis of $E^6$). | Half height of projected ASs in $\Delta_z$ ($E^6$ *unit*) | | | Step height unit $\delta$ ($E^6$ *unit*) | Step height unit $\delta$ in nm. AlPdMn (AlCuFe) |
|---|---|---|---|---|---|---|
| | | $n$ | $n'$ | $bc$ | | |
| $5f$ | $\|z_1\rangle = (5,\bar{1},\bar{1},\bar{1},\bar{1},1)/5$<br>$z\tau = (0,2,2,2,2,\bar{2})/5$ | $\frac{1+\tau}{\sqrt{2}}$ | $\frac{3(2+\tau)}{5\sqrt{2}}$ | $\frac{1}{\sqrt{2}}$ | $\delta_5 = \frac{\sqrt{2}(3-\tau)}{5}$<br>($\cong$ 0.3909) | 0.2521<br>(0.2468) |
| $3f$ | $z_1 = (1,1,1,\bar{1},\bar{1},\bar{1})/3$<br>$z\tau = (1,1,1,1,1,1)/3$ | $\sqrt{\frac{3(3+4\tau)}{10}}$ | $\sqrt{\frac{3(3+4\tau)}{10}}$ | $\sqrt{\frac{3(3-\tau)}{10}}$ | $\delta_3 = \sqrt{\frac{2(3-\tau)}{15}}$<br>($\cong$ 0.4293) | 0.2769<br>(0.2710) |
| $2f$ | $z_1 = (0,0,1,0,1,0)/2$<br>$z\tau = (1,1,0,0,0,0)/2$ | $\sqrt{\frac{7+11\tau}{10}}$ | $\sqrt{\frac{7+11\tau}{10}}$ | $\sqrt{\frac{2+\tau}{10}}$ | $\delta_2 = \sqrt{\frac{3-\tau}{10}}$<br>($\cong$ 0.3717) | 0.2398<br>(0.2347) |

**Table II: Basis of the $\Delta_z$ lattice, height of ASs and associated step height unit (projection of $|z_1\rangle$ on $E_{//}$) for the three main orientations.**

| $z$ axis | QC plane density along $z$ axis ( $plane\,/\,E6\ unit$ ) | | | |
|---|---|---|---|---|
| 5f | $n$-AS | $n'$-AS | $bc$-AS | Total |
| $5f$ | $\frac{1+3\tau}{\sqrt{2}}$ | $\frac{3\tau}{\sqrt{2}}$ | $\frac{-1+2\tau}{\sqrt{2}}$ | $4\sqrt{2}\ \tau$ |
| $3f$ | $2\sqrt{\frac{3}{10}(3+4\tau)}$ | $2\sqrt{\frac{3}{10}(3+4\tau)}$ | $3\sqrt{\frac{3}{10}(3-\tau)}$ | $3\sqrt{\frac{3}{10}(23+19\tau)}$ |
| $2f$ | $4\sqrt{\frac{7+11\tau}{10}}$ | | | $4\sqrt{\frac{7+11\tau}{10}}$ |

**Table III: Density of QC plane along the z axis $\rho_{AS}^{z} = h_{AS} / S_{\Delta z}$, see Appendix.**

| Distance in $E_{//}$ between AS (step heights) | Step height ($\delta_n$ unit) | Step height (nm for i-AlPdMn) | Step height probability |
|---|---|---|---|
| | | $5f$ | |
| $n-bc$ | $-\frac{3}{2}+\tau$ | 0.0298 | $\frac{2-\tau}{4}$ |
| $n-n'$ | $1-\tau/2$ | 0.0481 | $1/2$ |
| $bc-n'$ | $\frac{-1+\tau}{2}$ | 0.0779 | $\frac{1}{4}$ |
| $n-n$ | $-1+\tau$ | 0.1558 | $\frac{-1+\tau}{4}$ |
| | | $3f$ | |
| $n-n'$ | $\frac{5-3\tau}{2}$ | 0.0202 | $\frac{13-3\tau}{33}$ |
| $n'-bc$ | $\frac{-3+2\tau}{2}$ | 0.0327 | $\frac{4(5-2\tau)}{33}$ |
| $n-bc$ | $1-\frac{\tau}{2}$ | 0.0529 | $\frac{2(5-2\tau)}{33}$ |
| $n-n$ | $-3+2\tau$ | 0.0654 | $\frac{5-2\tau}{33}$ |
| $n-n'$ | $\frac{-1+\tau}{2}$ | 0.0856 | $\frac{-15+17\tau}{33}$ |
| | | $2f$ | |
| $(n,n)-(n,n')$ | $-3+2\tau$ | 0.0566 | $1-\tau/2$ |
| $(n,n)-(n,n')$ $n-bc$ | $2-\tau$ | 0.0916 | $1/2$ |
| $(n,n)-(n,n')$ $n-bc$ | $-1+\tau$ | 0.148 | $\frac{-1+\tau}{2}$ |

**Table IV: Smallest step heights (given by the smallest distances along $\hat{z}_{//}$ between facing ASs in $\Delta_z$) and associated probabilities (given by their facing height along the $E_\perp$ axis of $\Delta_z$).**

| $\langle\alpha\|i\rangle$ or $\langle\beta\|i\rangle$ | | $\|i\rangle$ | | | | | |
|---|---|---|---|---|---|---|---|
| | | (1,0,0,0,0,0) | (0,1,0,0,0,0) | (0,0,1,0,0,0) | (0,0,0,1,0,0) | (0,0,0,0,1,0) | (0,0,0,0,0,1) |
| $\|\alpha\rangle$ | 1 | 1 | $\tau$ | 0 | -1 | $\tau$ | 0 |
| | 2 | $\tau$ | 0 | 1 | $\tau$ | 0 | -1 |
| | 3 | 0 | 1 | $\tau$ | 0 | -1 | $\tau$ |
| $\|\beta\rangle$ | 1 | $-\tau$ | 1 | 0 | $\tau$ | 1 | 0 |
| | 2 | 1 | 0 | $-\tau$ | 1 | 0 | $\tau$ |
| | 3 | 0 | $-\tau$ | 1 | 0 | $\tau$ | 1 |

**Table A-I: Coordinates of the basis vector ($|i\rangle$) of $E^6$ in the $\{|\alpha\rangle, |\beta\rangle\}$ basis of ($E_{//}, E_{\perp}$) as defined by Cahn, Shechtman and Gratias (CSG notations). All coefficients must be multiplied by the normalization factor:** $K = 1/\sqrt{2(2+\tau)} = \sqrt{(3-\tau)/10}$ **.**

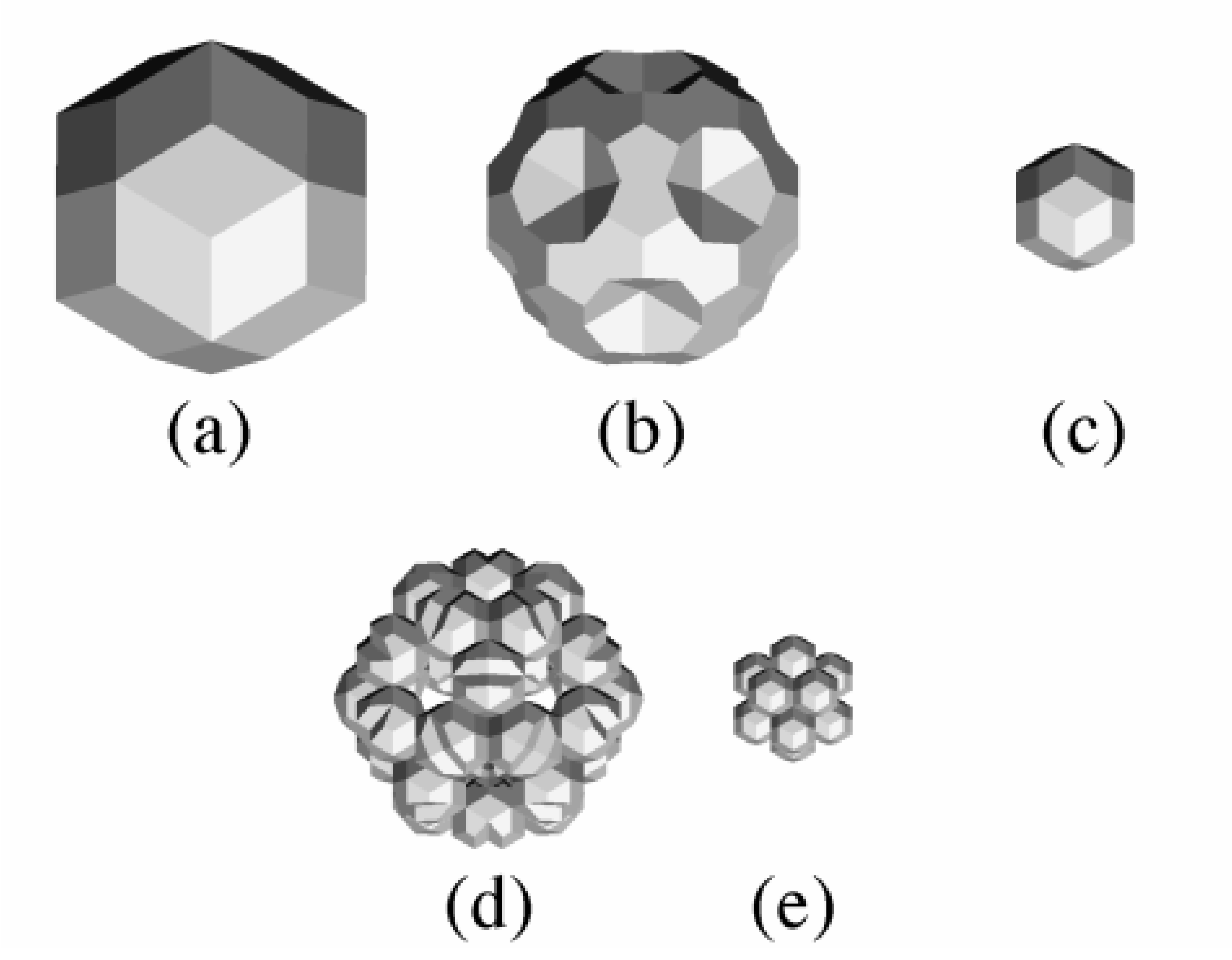

**Figure 1: Simple model for *i*-AlPdMn. (a-c) The 3 atomic surfaces (ASs) attached to $n$, $n'$ and $bc$ nodes. (a) the $n$-AS is a rhombic triacontahedron, (b) n'AS is the same rhombic triacontahedron but truncated along the $5f$ symmetry axes. (c) The height of the $bc$-AS is $\tau^2$ times smaller than that of the $n$-AS. The $n'$-AS surfaces decomposes in two assemblies of small triacontahedra Mn (d) and Pd (e), the remaining space is Al filled.**

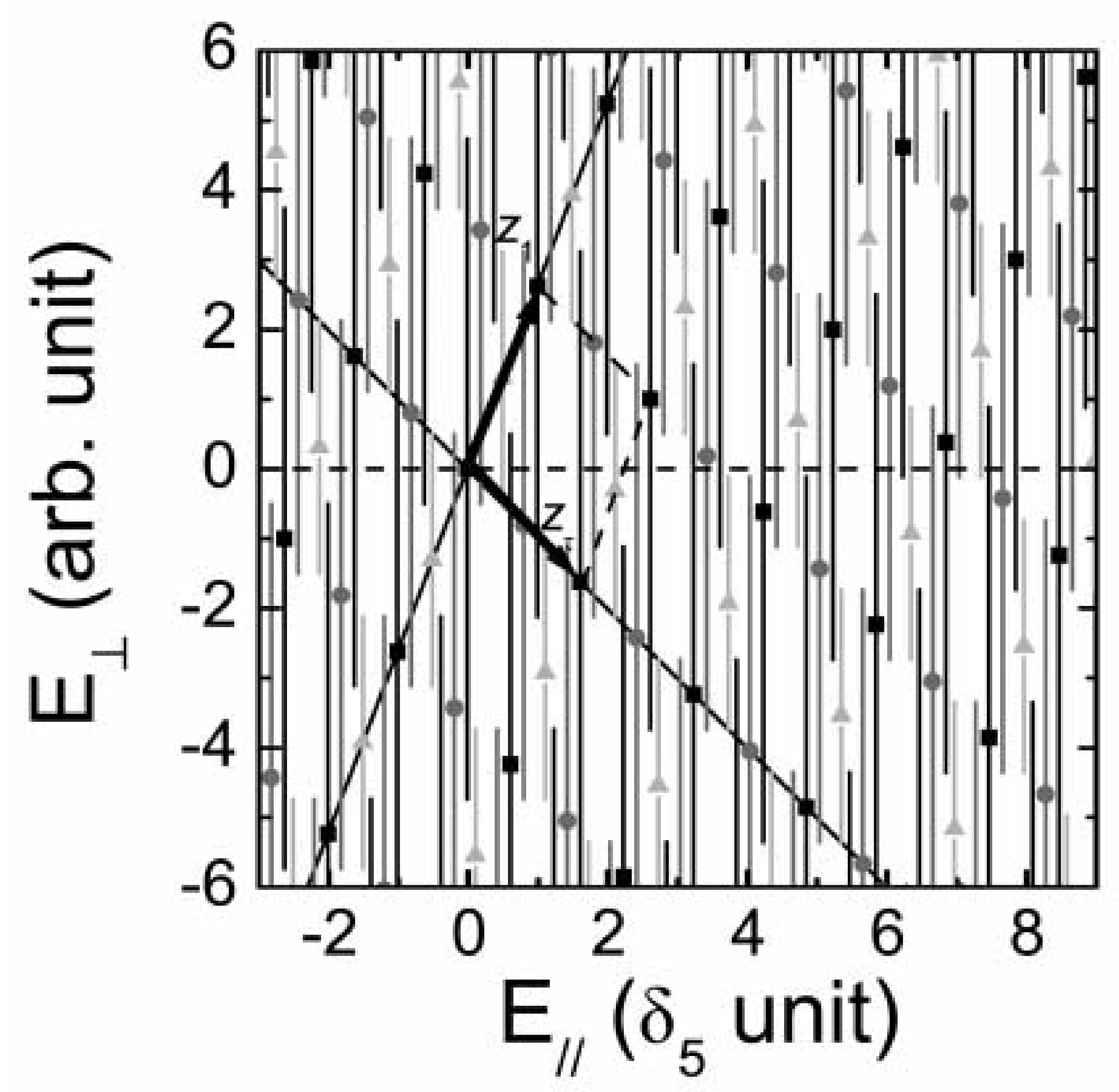

**Figure 2: *i*-AlPdMn.** $\Delta_z(5f)$, **projected lattice decorated by ASs for the** $5f$ **orientation.** $\{z_1, z_\tau\}$ **is the chosen basis of** $\Delta_z$ **which coordinates in the** $\{|i\rangle\}$ **basis of** $E^6$ **are given in Appendix 1.** $\Delta_z$ **nodes: (■)**$n$ **nodes, (●)**$n'$ **nodes and (▲)**$bc$ **nodes. Vertical bars associated to each node (same color): projection in** $\Delta_z$ **of the convex envelope of associated AS. Dashed line: line of cut within** $E_{//}$ **along the z axis perpendicular to the terrace planes.**

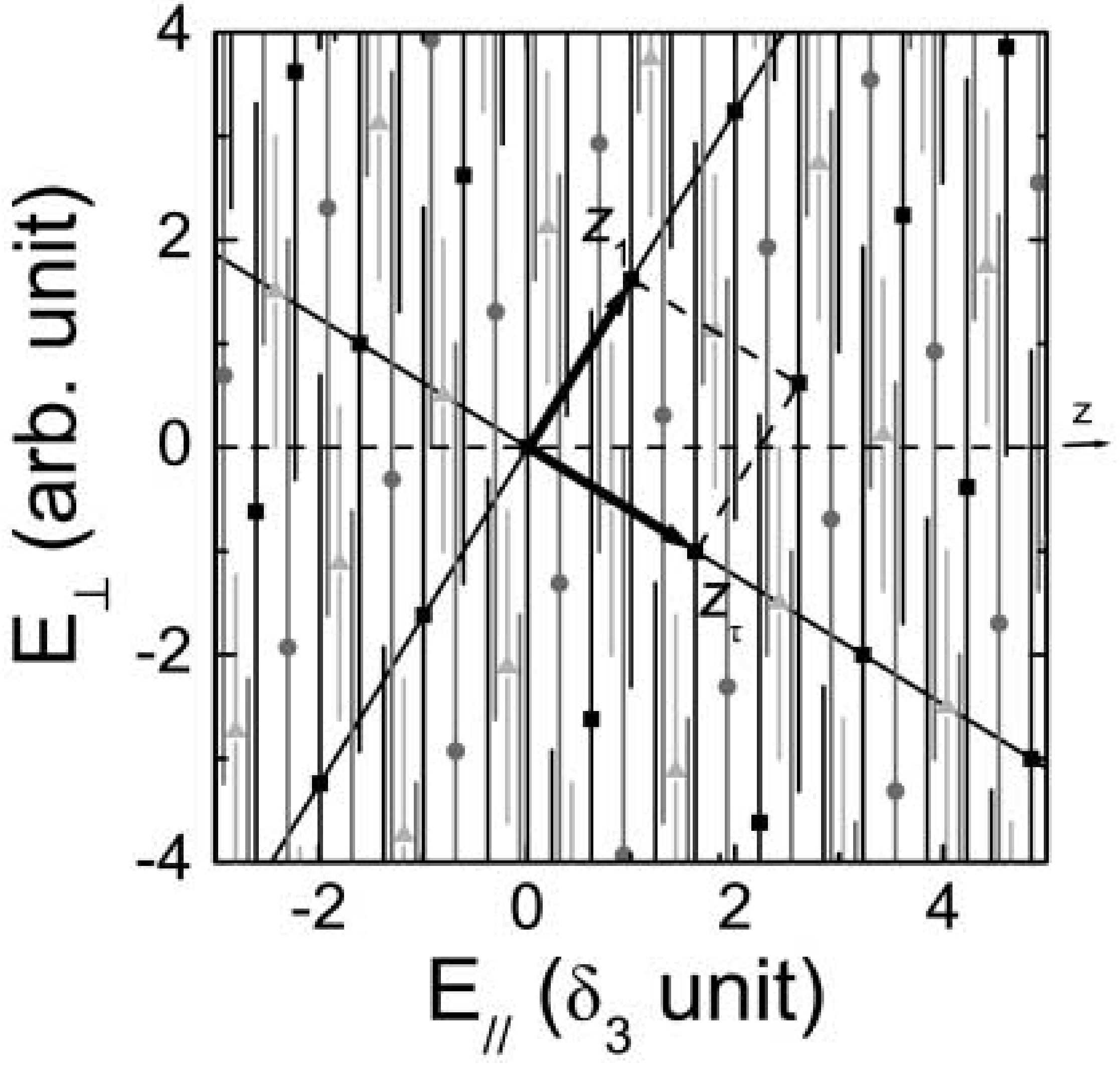

**Figure 3:** $\Delta_z(3f)$**, same as Figure 2 for the** $3f$ **orientation.**

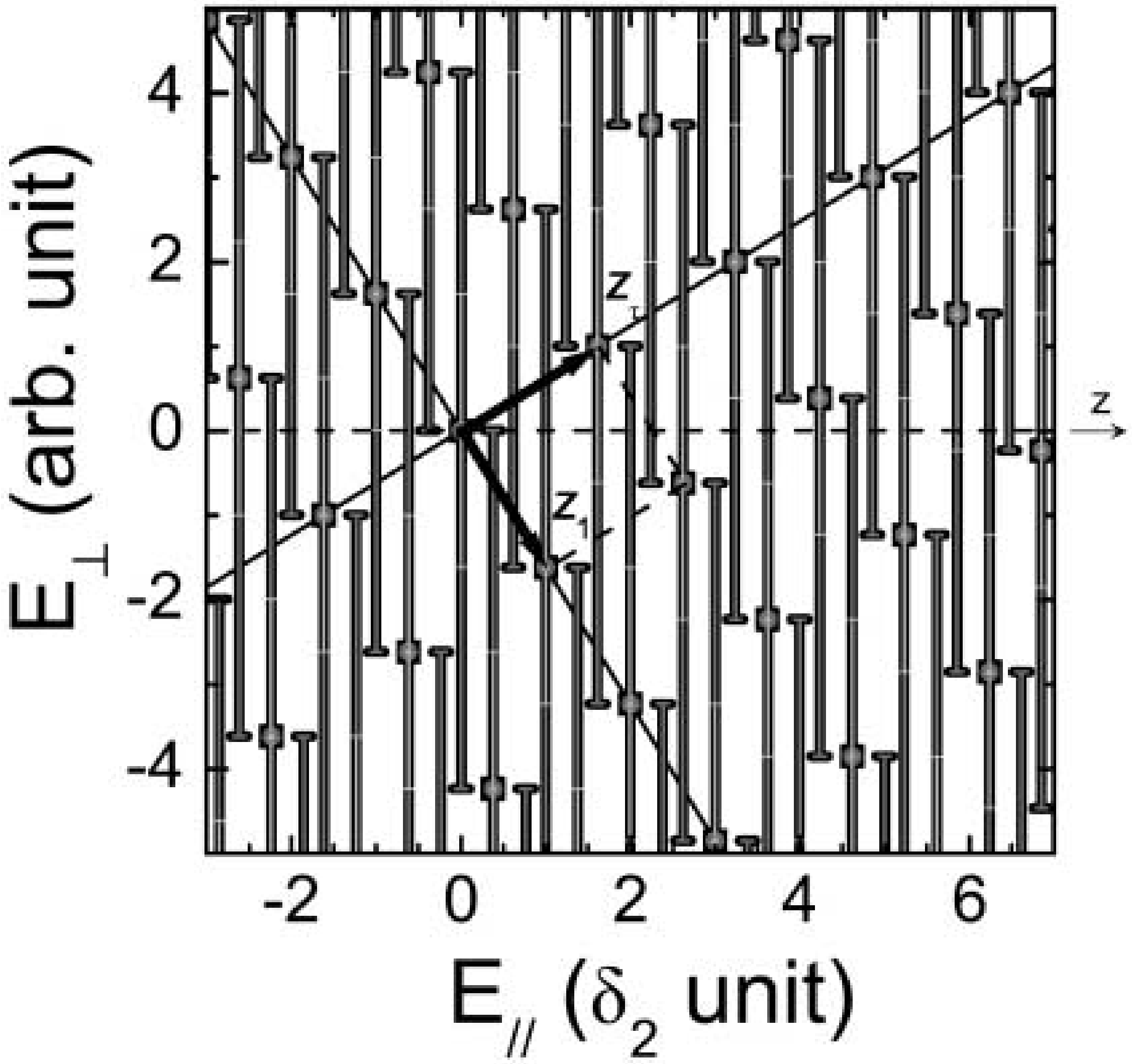

**Figure 4: Same as Figure 2 for the $2f$ orientation. : Projection of (■) $n$, (●) $n'$ and (▲) $bc$ nodes are superimposed. Projection of $n$-As and $n'$-AS are of the same size.**

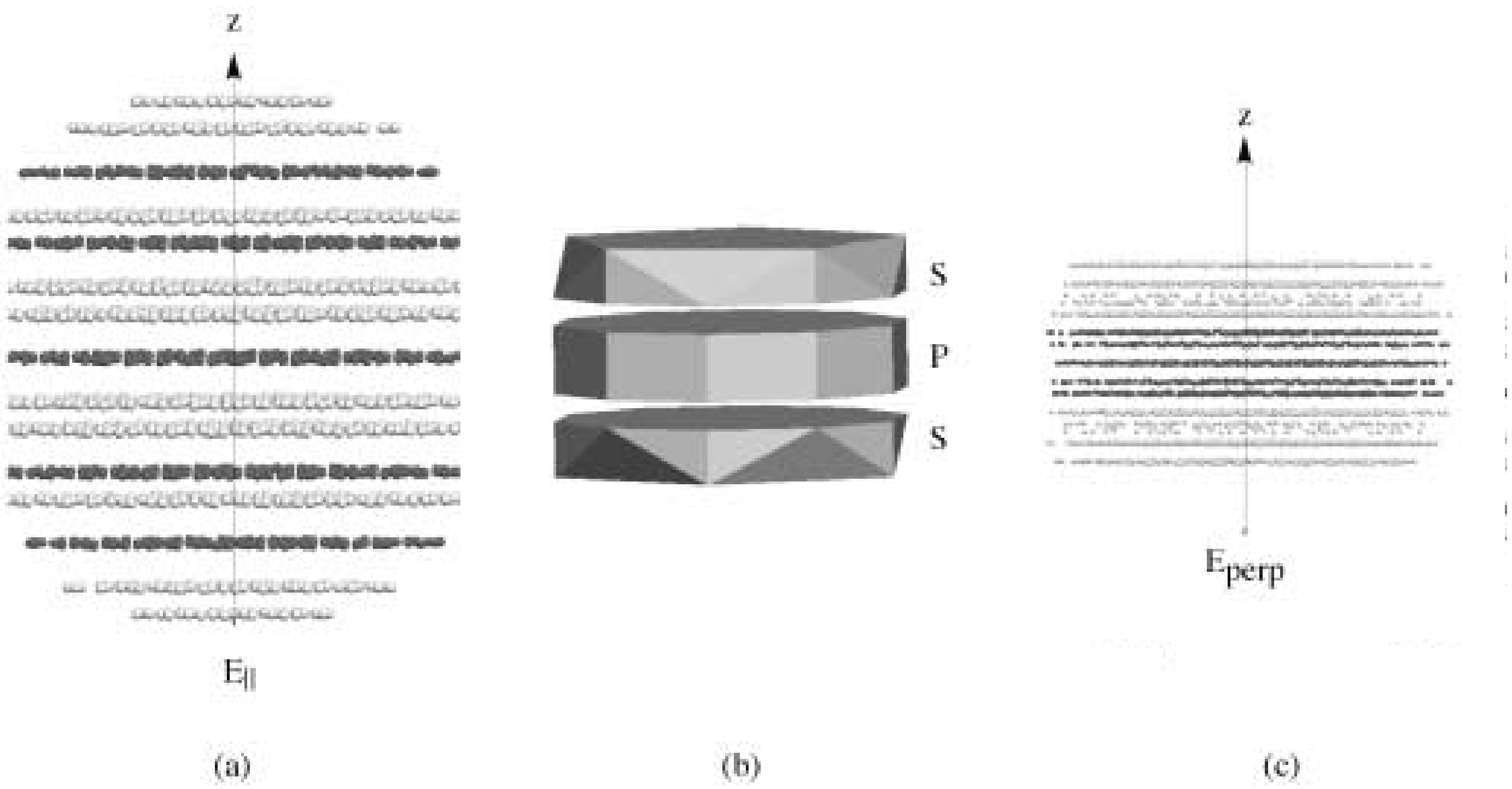

**Figure 5: (a) Stacking sequence of densest planes along the 5-fold direction generated by the acceptance windows shown on (b). Planes issued from the central decagonal slice (P) are in dark grey and from the upper and lower slices (S) in light grey. (c) Their image in $E_\perp$ follow the cut algorithm: dark-grey planes have their image inside the P window whereas the light-grey ones distribute in the S window.**

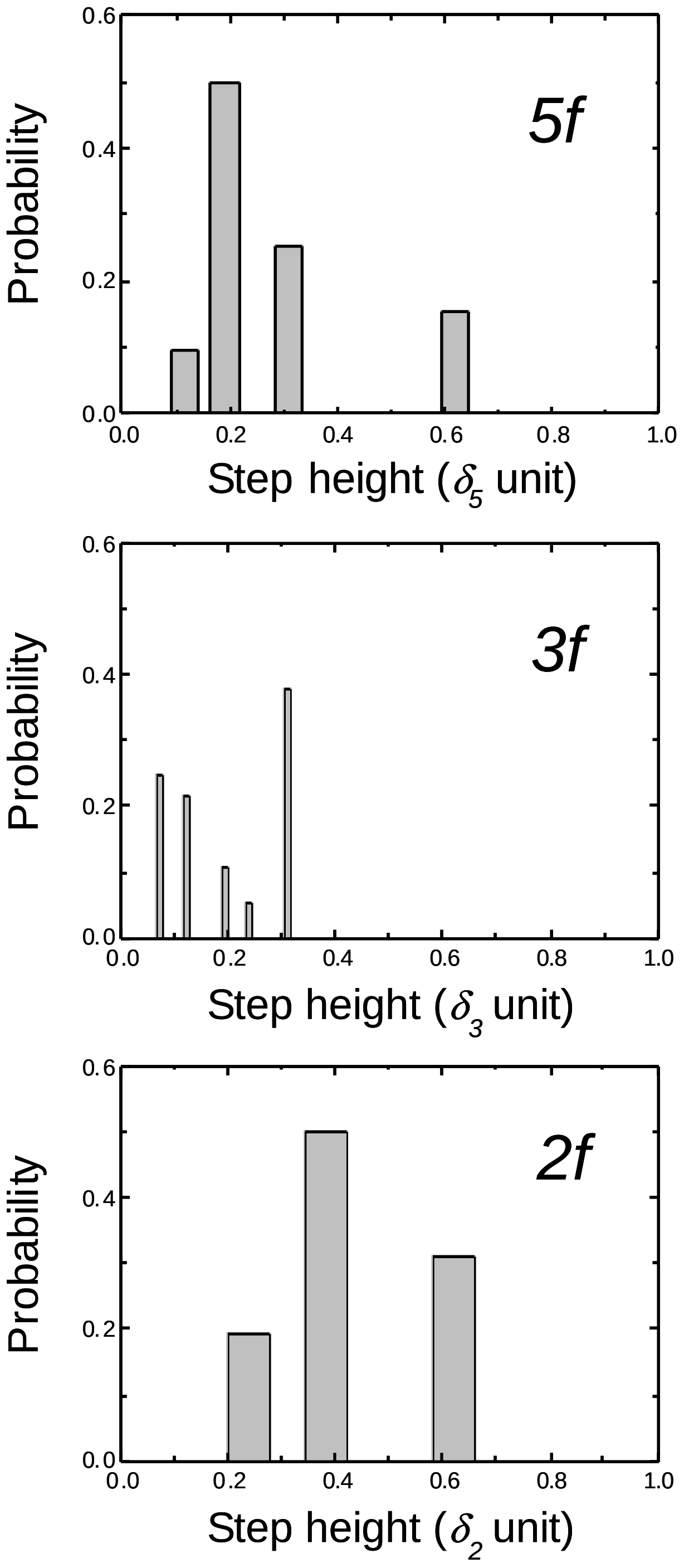

**Figure 6: Histograms of step heights for the three orientations. Step heights are given by the first distances between facing ASs. Probabilities are given by their facing height along the $E_\perp$ axis of $\Delta_z$.**

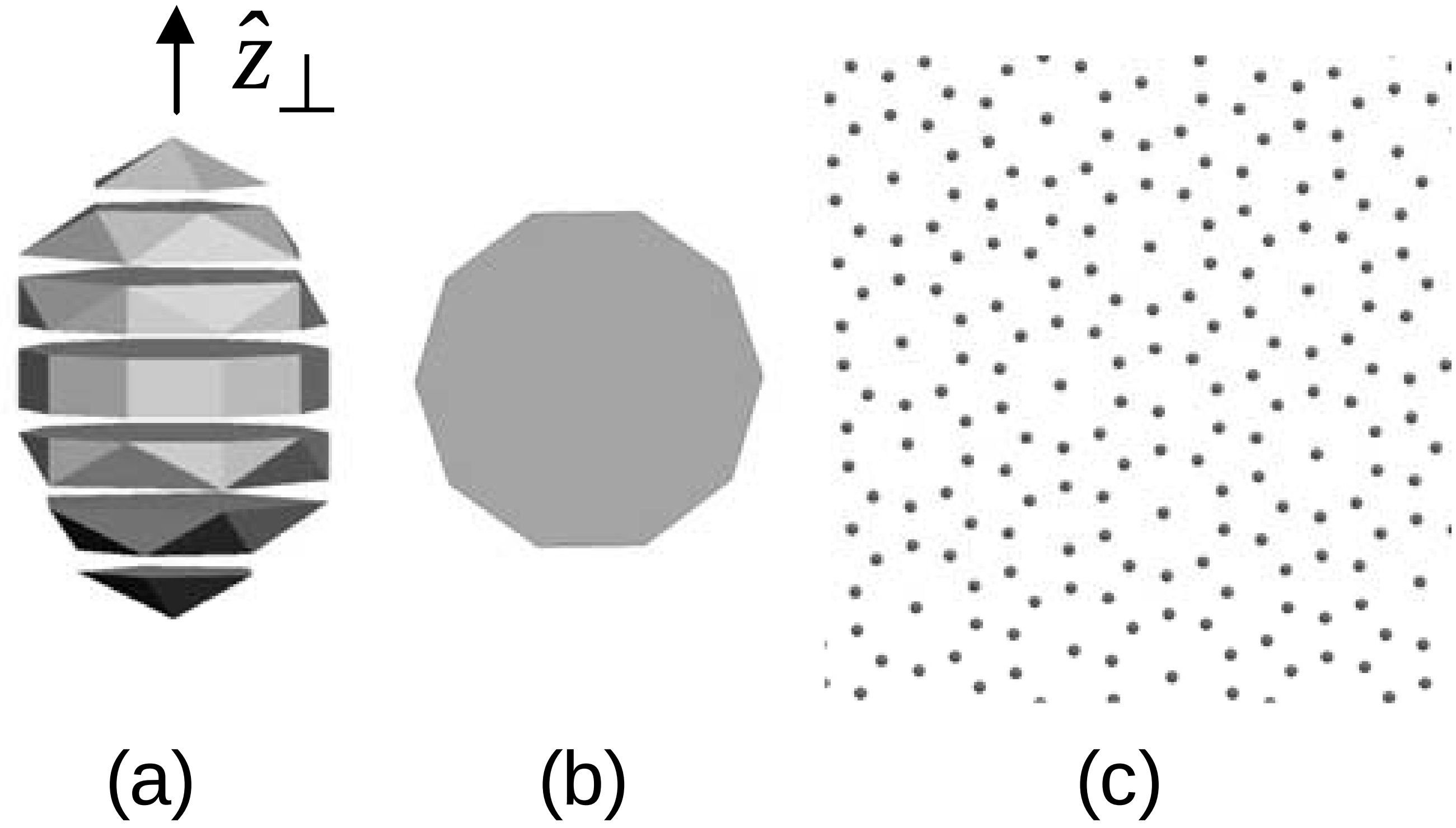

**Figure 7: a) Decomposition of the $n$-AS for the $5f$ orientation. Within each slice the in plane density decays quadratically with z (see Figure 8). b) Truncations within the central slice generate a decagonal AS of constant size. c) Typical locally 10 fold atomic configuration of a QC plane generated by a cut within this central slice. Truncation within upper and lower slices generates planes with local patterns of $5f$ symmetry (See Figure 9).**

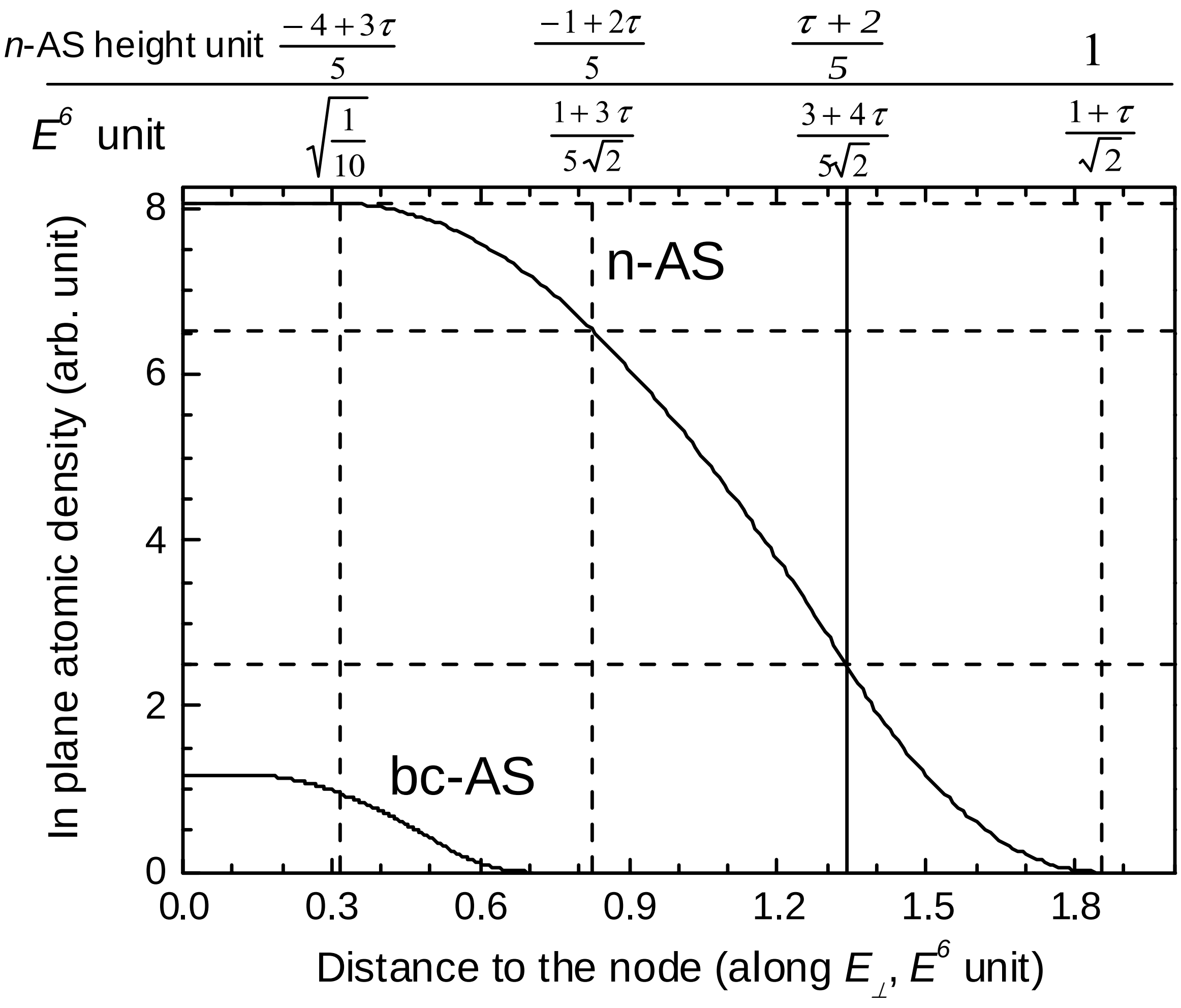

**Figure 8: In plane density versus the cut distance (along $\hat{z}_{\perp}$) to the node for $n$-AS, and $bc$-AS. Density for $n'$-AS (slightly truncated triacontahedron along the $5f$ direction) is very close to $n$-AS.**

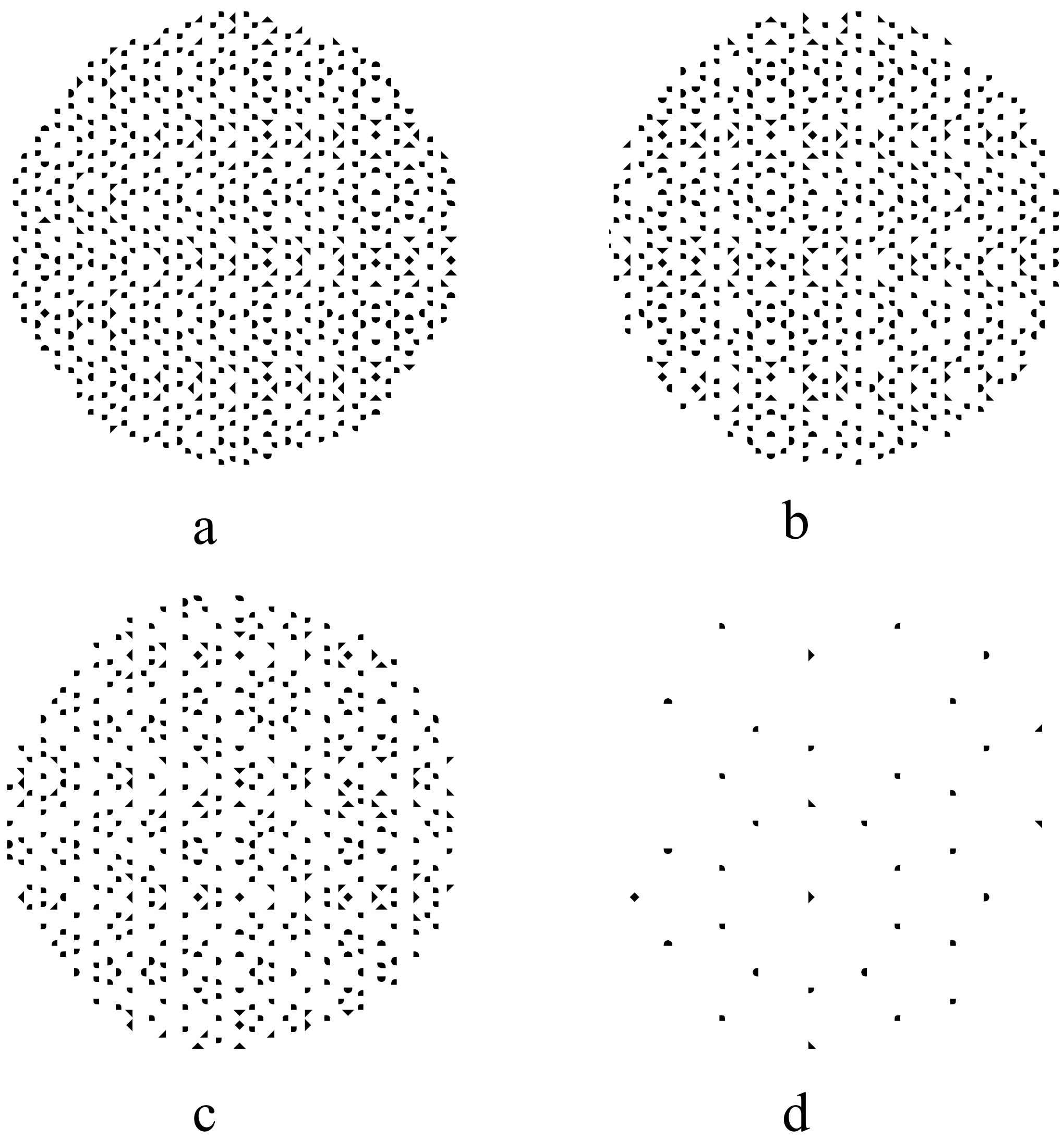

**Figure 9: In plane atomic configurations of terrace planes for a cut of $n$-AS in $\Delta_z(5f)$ within the different slices (distance to the $n$-node: $\delta z_\perp$). The atom density decays with the area of the $n$-AS at the cut level according to Figure 8. a) Cut within the central decagonal prism: 1st slice $(z,\delta z_\perp)=(0,0)\,\delta_5$, b) Cut within 2nd slice: $(z,\delta z_\perp)=(\tau,\tau)\,\delta_5$, c) Within 3rd slice: $(z,\delta z_\perp)=(1,-1-\tau)\,\delta_5$, d) Within 4th (upper) slice: $(z,\delta z_\perp)=(-1+\tau,1+2\tau)\,\delta_5$.**

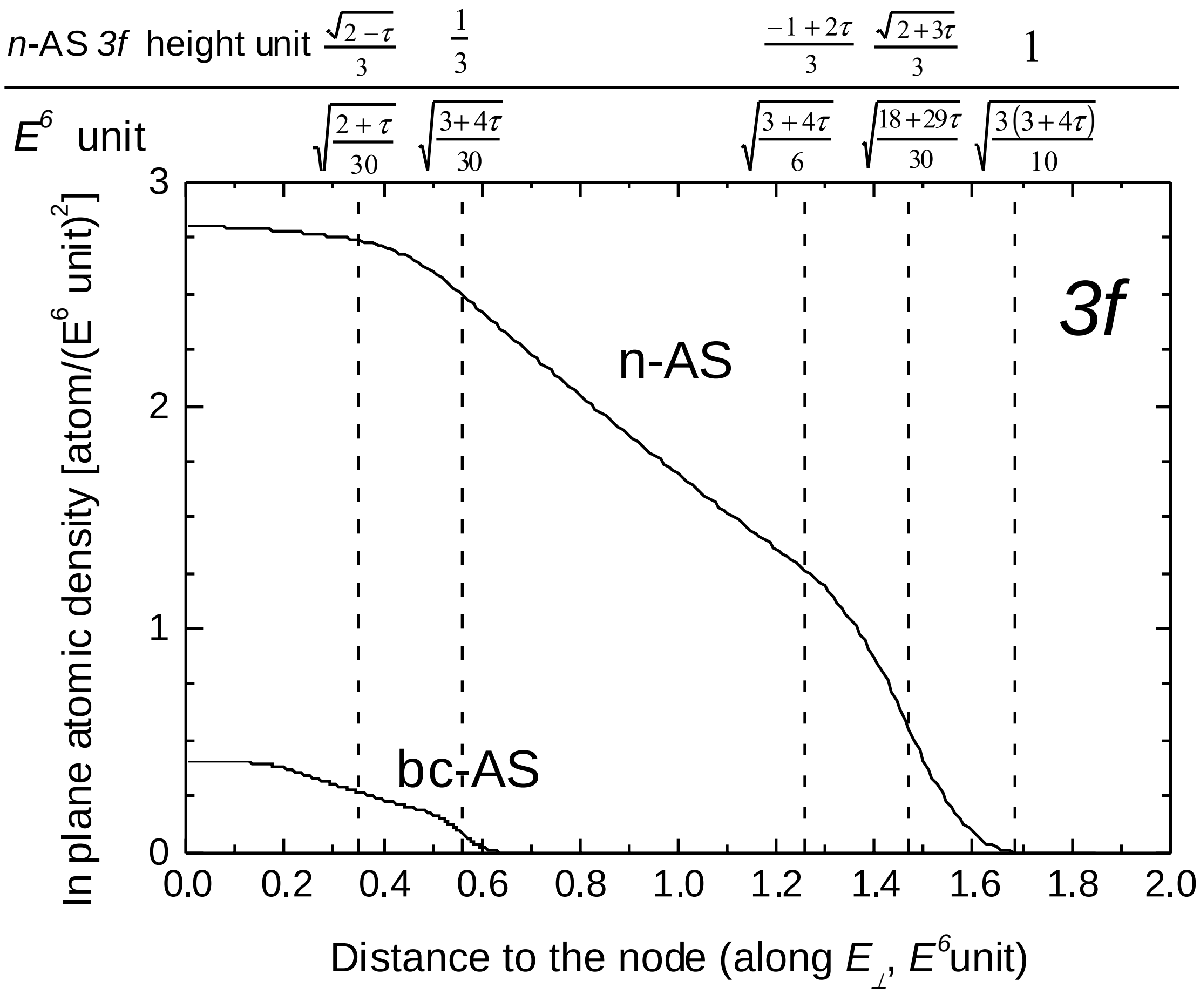
n-AS 3f height unit
$\frac{\sqrt{2-\tau}}{3}$ $\frac{1}{3}$ $\frac{-1+2\tau}{3}$ $\frac{\sqrt{2+3\tau}}{3}$ 1
$E^6$ unit
$\sqrt{\frac{2+\tau}{30}}$ $\sqrt{\frac{3+4\tau}{30}}$ $\sqrt{\frac{3+4\tau}{6}}$ $\sqrt{\frac{18+29\tau}{30}}$ $\sqrt{\frac{3(3+4\tau)}{10}}$
In plane atomic density [atom/(E$^6$ unit)$^2$]
3f
n-AS
bc-AS
Distance to the node (along $E_\perp$, $E^6$unit)
0
1
2
3
0.0
0.2
0.4
0.6
0.8
1.0
1.2
1.4
1.6
1.8
2.0

Figure 10

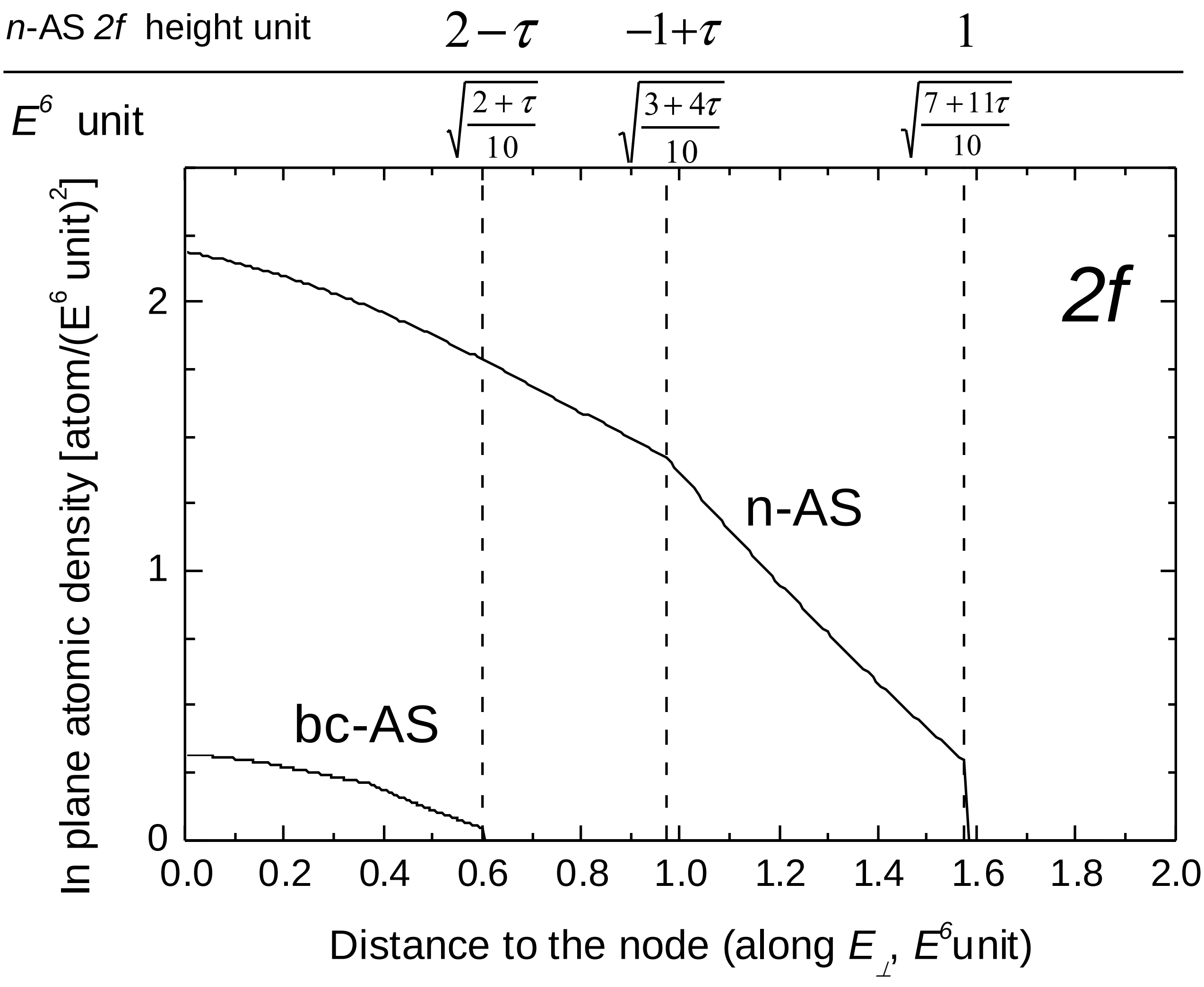

**Figure 10: In plane density for $n$-AS, and $bc$-AS along the $3f$ and $2f$ orientations versus the cut distance (along $\hat{z}_\perp$) to the node. Density for $n'$-AS is very close to $n$-AS.**

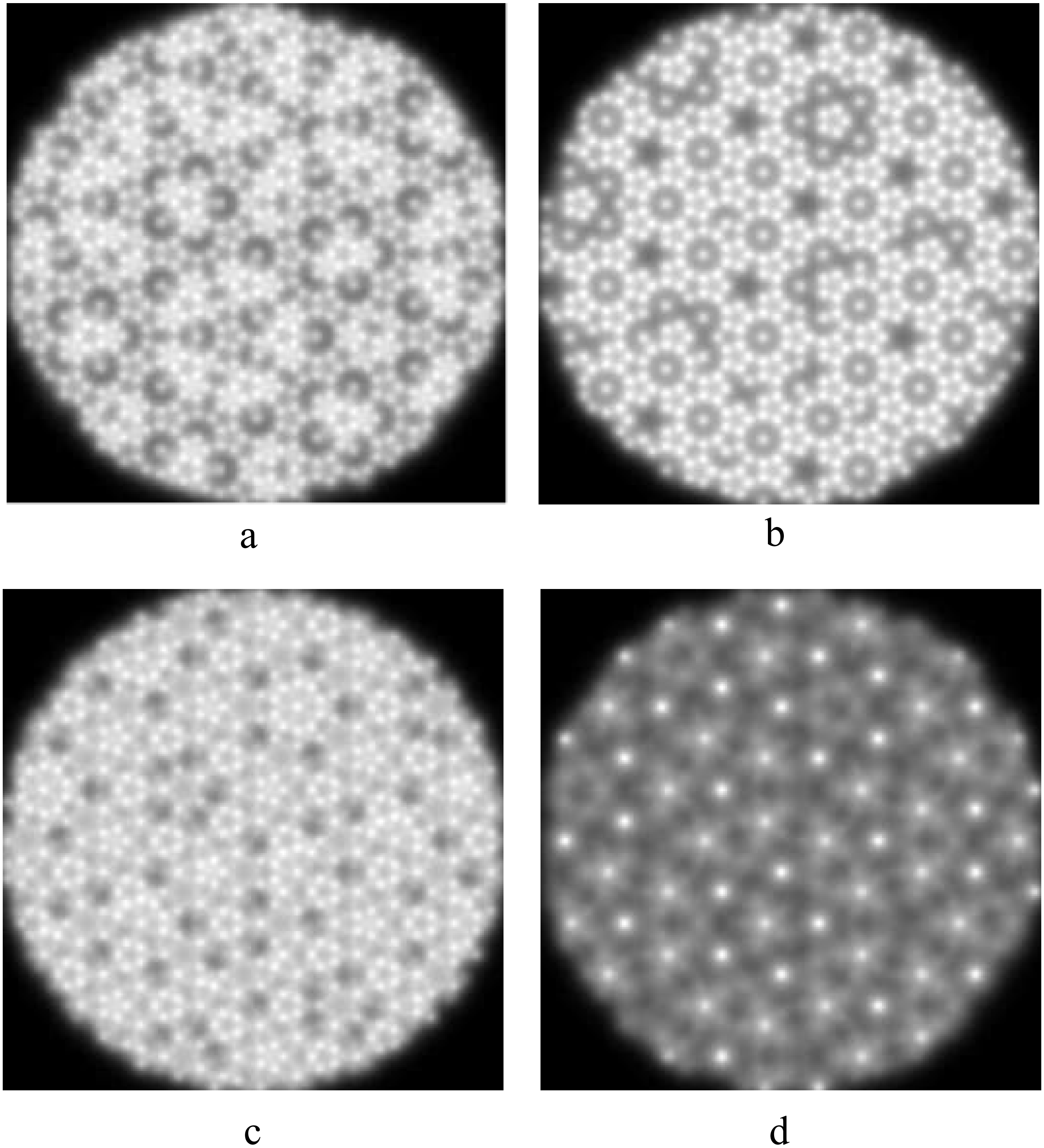

**Figure 11: Simulated STM images of 4 terrace planes generated by a cut at $z_{\perp} = 0$ in $\Delta_z(5f)$. $n$-AS (same as Figure 9) are cut at different heights from the $n$-node ($\delta z_{\perp}$). a) $(z, \delta z_{\perp}) = (0,0)\,\delta_5$. b) $(z, \delta z_{\perp}) = (\tau, \tau)\,\delta_5$. c) $(z, \delta z_{\perp}) = (\tau, \tau)\,\delta_5$, the $n$-AS plane is**

**just above a $n'$-AS cut within the central slice: a smooth dense surface is obtained. d) $(z,\delta z_{\perp})=(-1+\tau,1+2\tau)\delta_5$, no dense plane is present below the terrace, and deep depressions / high protrusions are seen.**

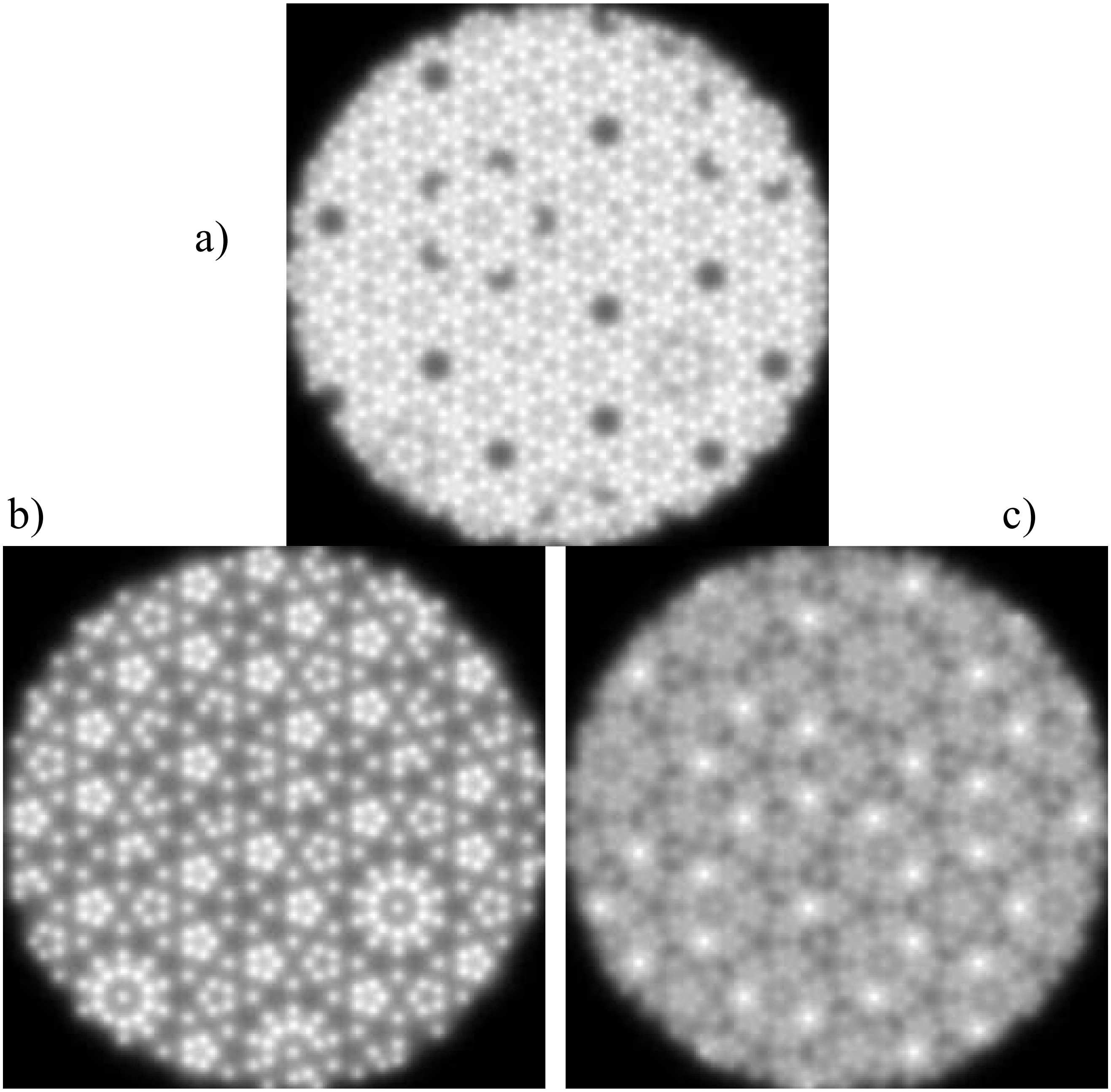

**Figure 12: Simulated STM images of terrace planes generated by a cut at $z_{\perp} = 0$ in $\Delta_Z(5f)$. a) 2nd slice $(z, \delta z_{\perp}) = (-\tau, -\tau)\delta_5$. b) 3rd slice $(-1, 1+\tau)\delta_5$ and c) 4th slice $(1-\tau, -1-2\tau)\delta_5$. $n$-AS are cut at the same distance to the node than in Figure 11 (respectively b, c and d) and the difference in contrast comes from the density of the underneath plane.**

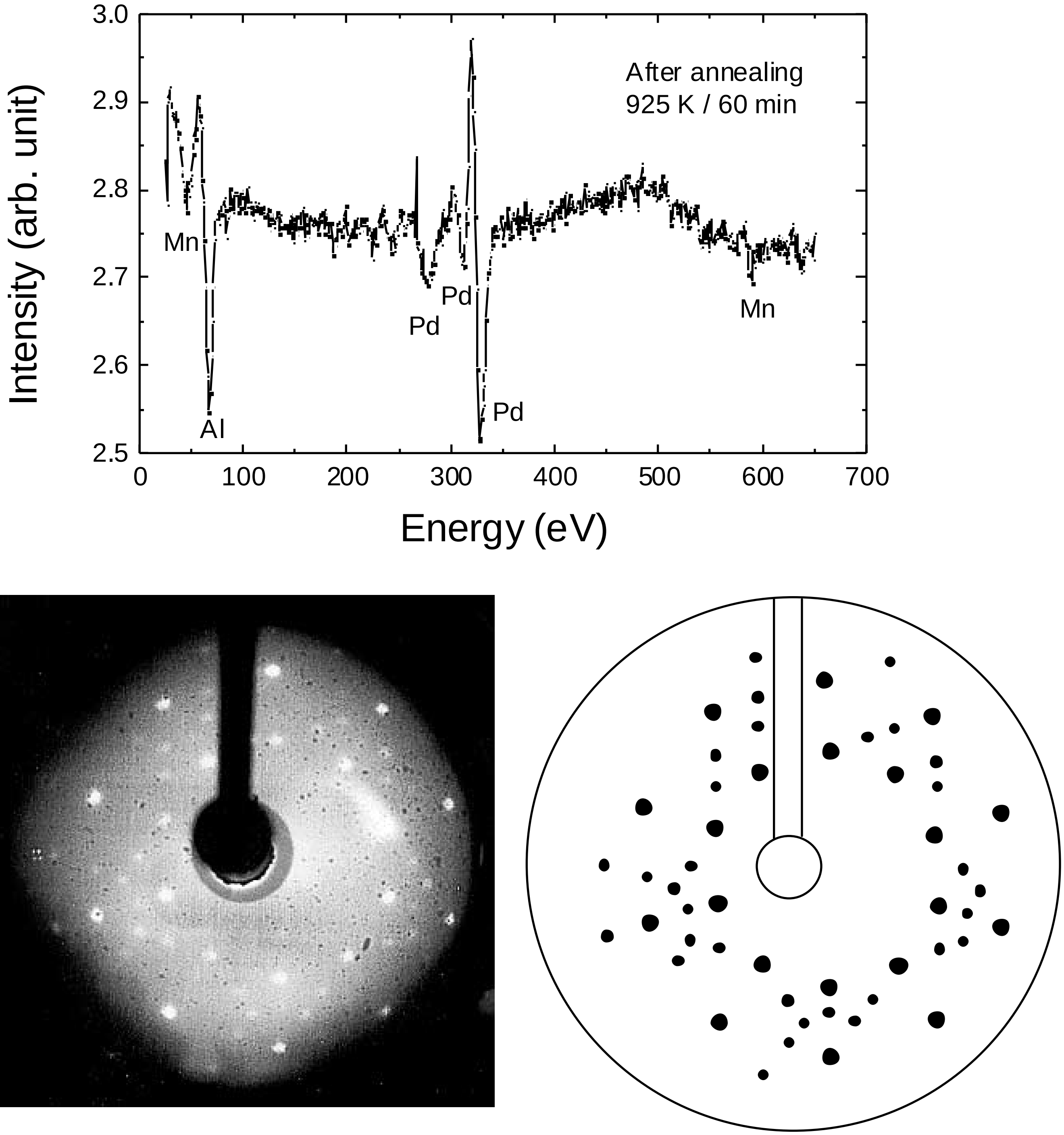

**Figure 13: Typical AES spectrum and LEED diffraction pattern after 1 h $Ar^+$ sputtering, and 925 K, 1 h annealing.**

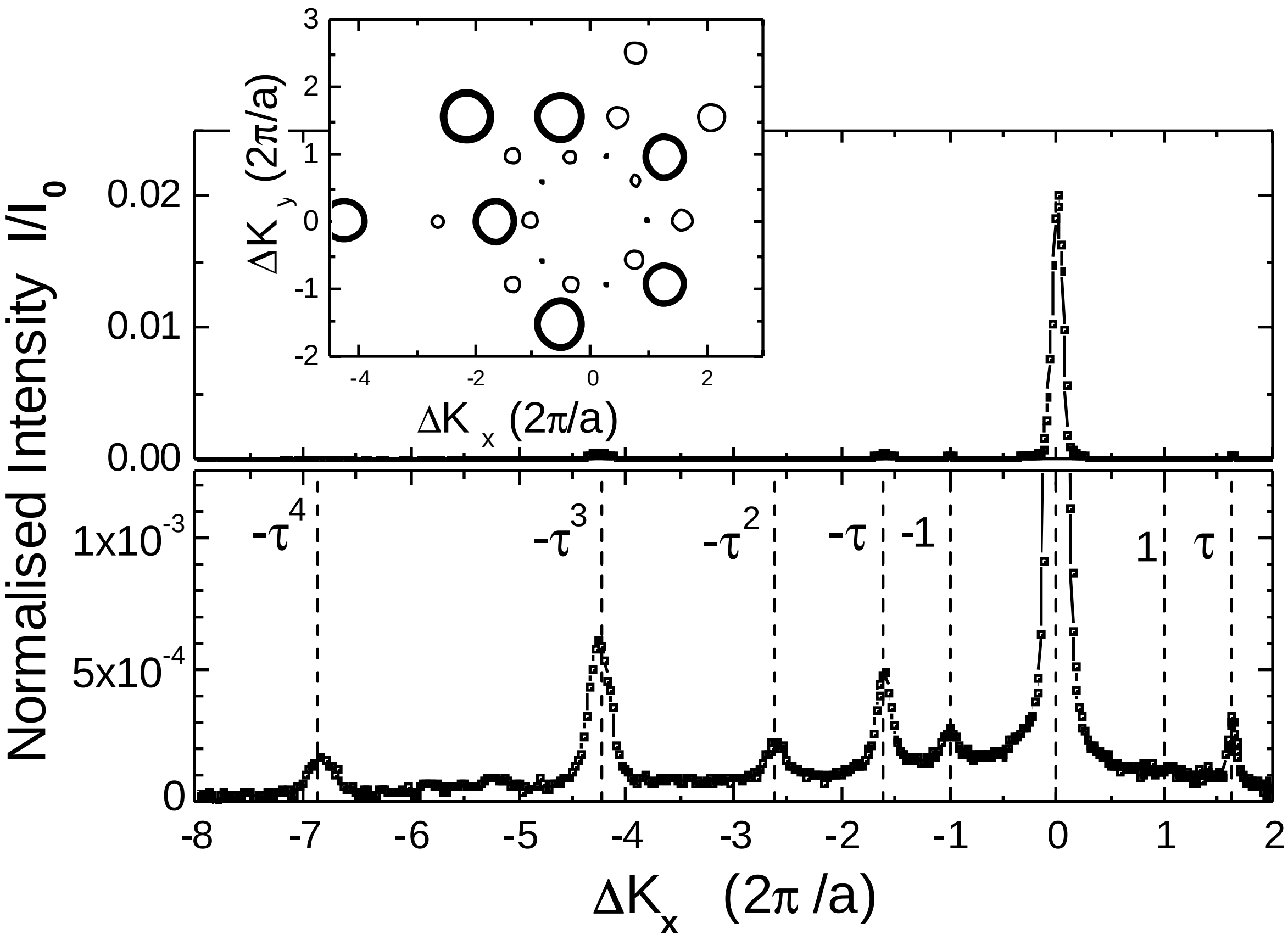

**Figure 14: Typical high resolution He diffraction spectrum at room temperature (wave number of the incident beam: $K_i$ = 65 nm$^{-1}$ incidence angle: $\theta_i$ = 58°, $a$ = 1.7 nm), $\Delta K_x$ designates the momentum transfer in the incident plane. Top: normalized intensity (not corrected for Debye-Waller attenuation). Bottom: intensity scale (× 20). Diffraction peaks ($G$,0) are observed up to the 5$^{th}$ order. Inset: fivefold diffraction pattern. Circles are proportional to peak intensities.**

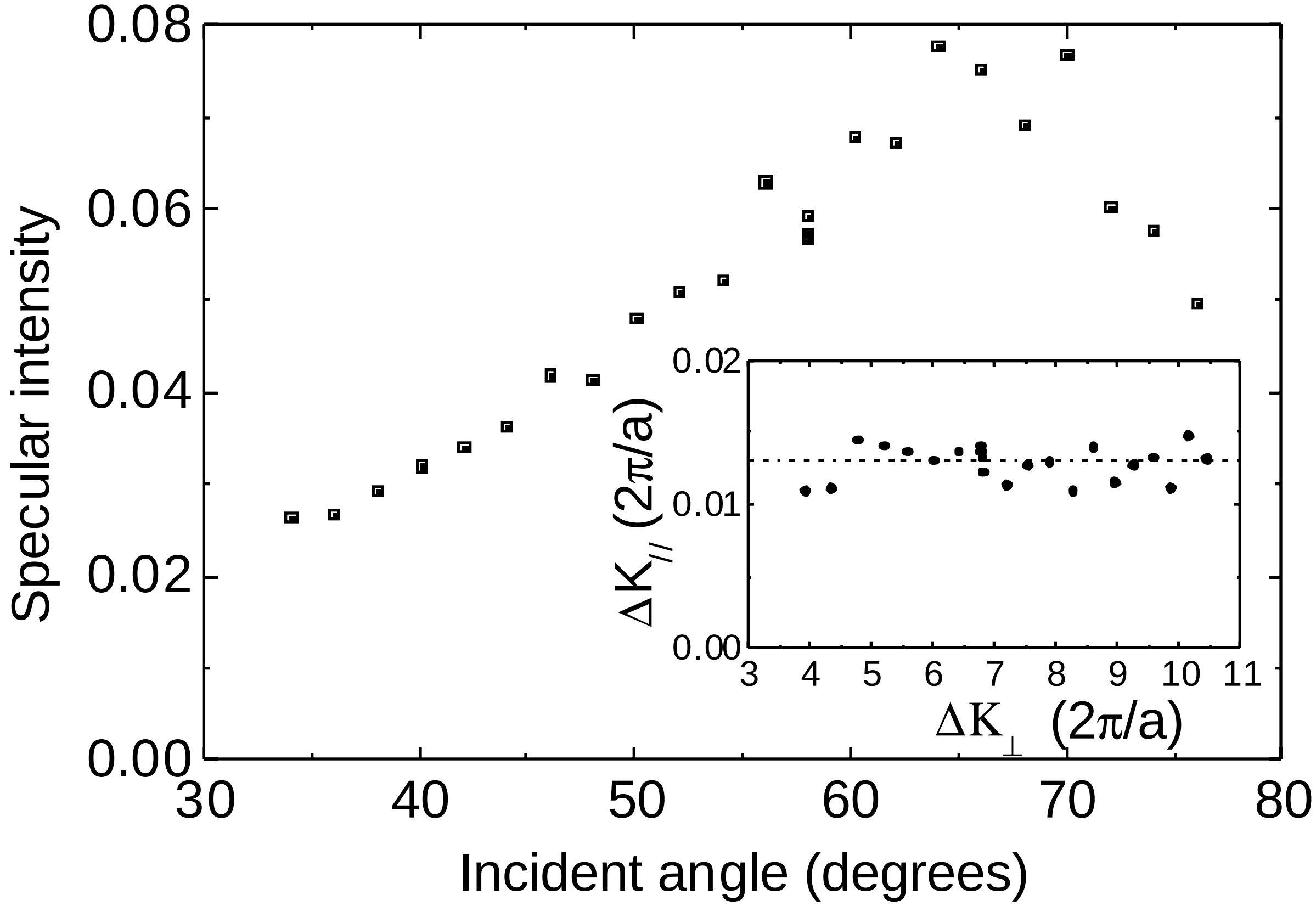

**Figure 15: He reflectivity (intensity of the specular peak ($I_0$)/incident beam intensity) versus incident angle $\theta_i$ with respect to the surface normal. Inset: Full width at half maximum of the specular peak (reciprocal space units and after subtraction of the instrument contribution) versus perpendicular momentum transfer $\Delta K_\perp$.**

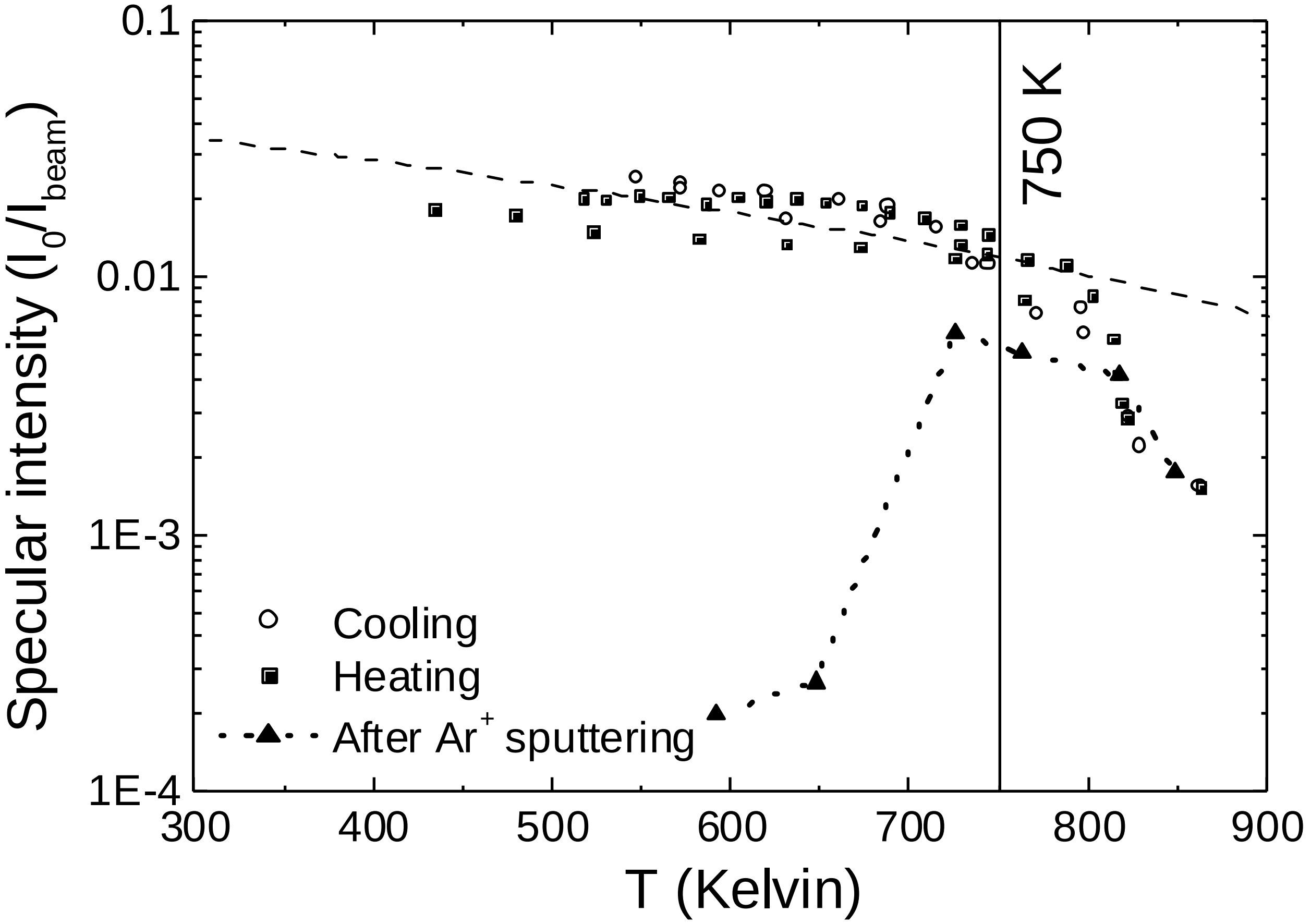

**Figure 16: Intensity of the specular diffraction peak (He reflectivity) versus temperature. Dashed line: Normalized Debye Waller factor for Cu surfaces (including anharmonicity) showing a typical thermal attenuation for He diffraction on dense surfaces[47 48]. Dotted line: Intensity of the specular peak versus temperature after $Ar^+$ bombardment during subsequent annealing.**

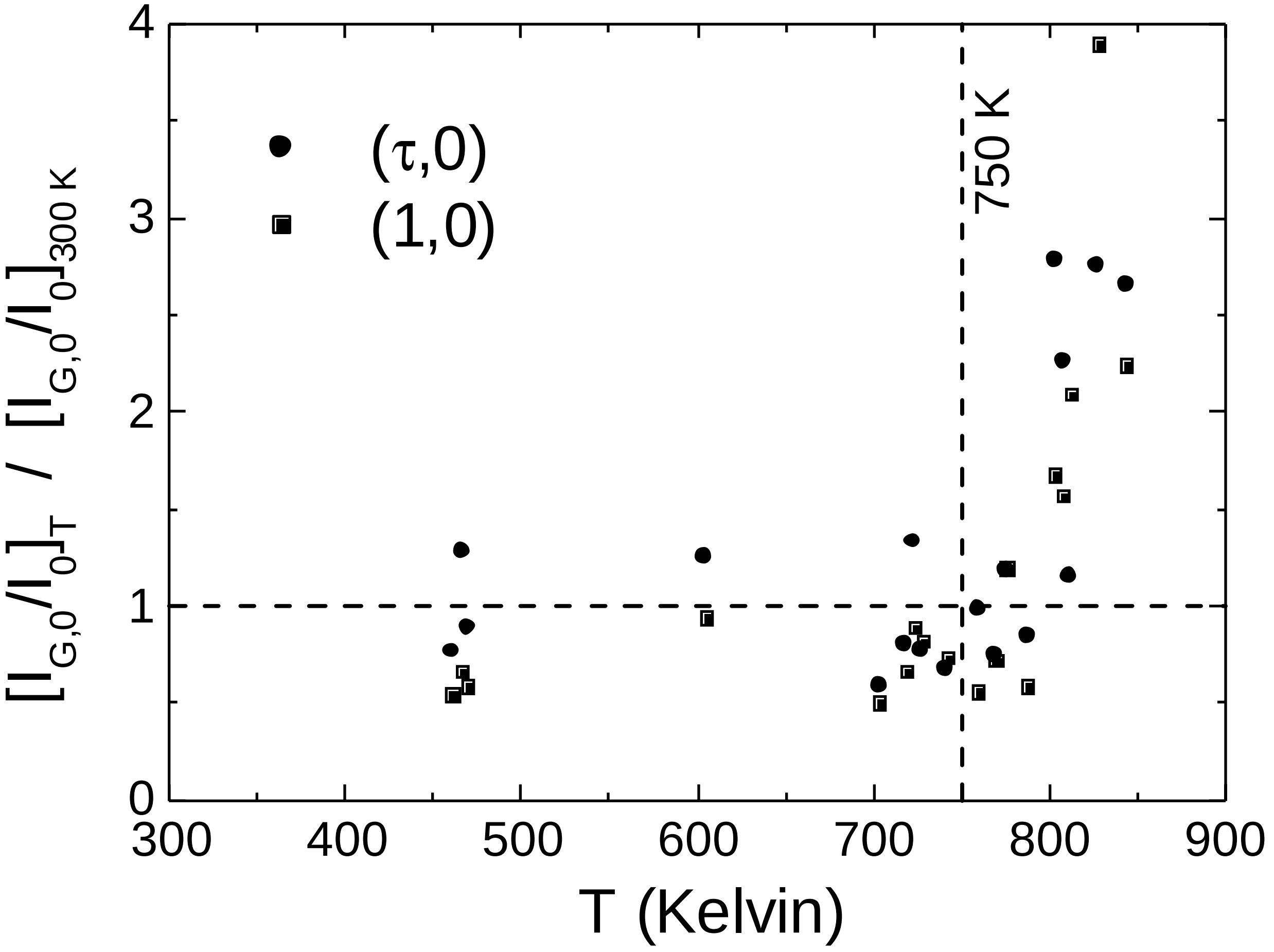

**Figure 17: Relative intensity of He diffraction peaks adjacent to the specular. Above 750 K a rapid increase of their relative intensity is observed.**

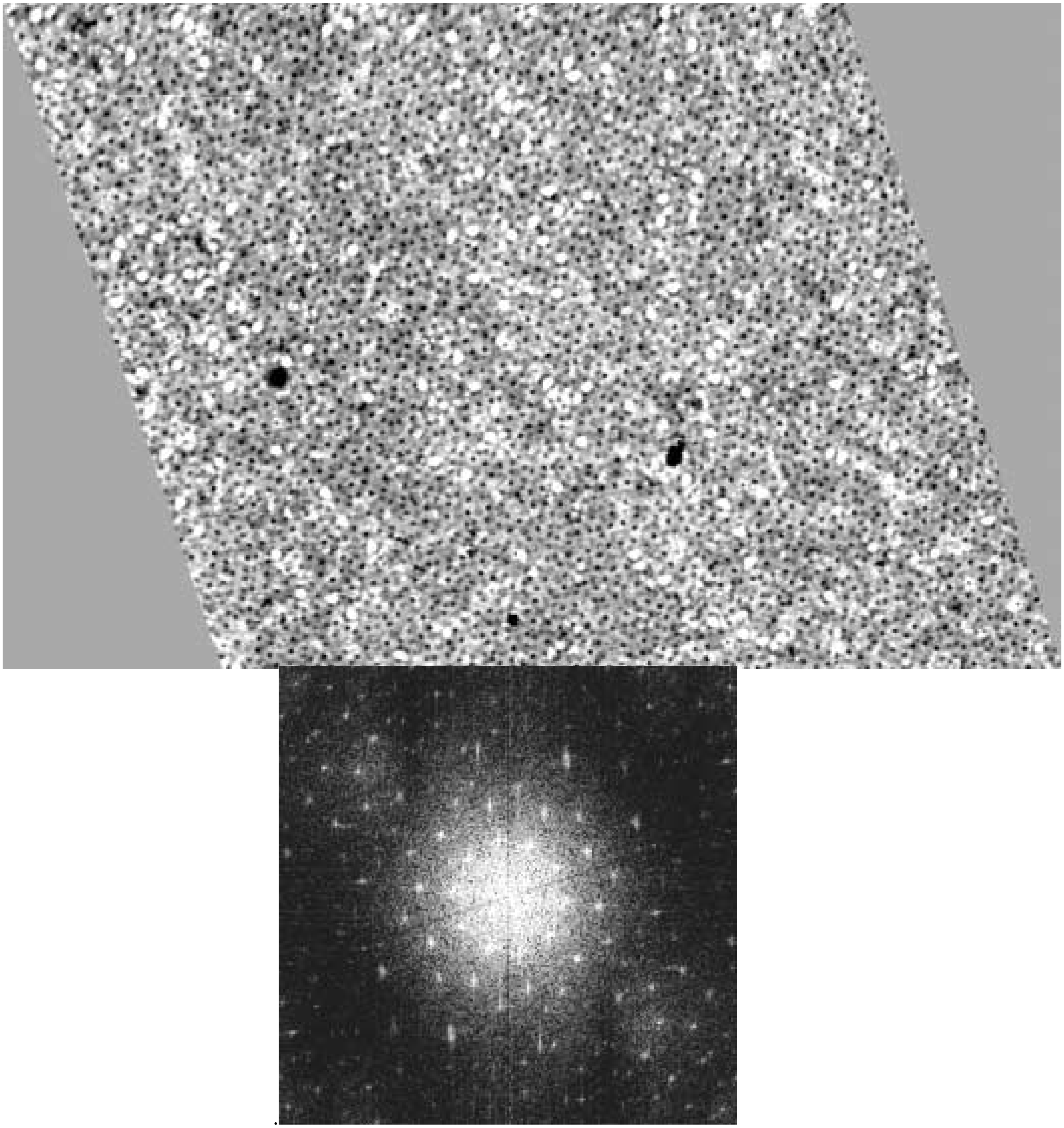

**Figure 18: $96 \times 76$ nm$^2$ STM image of a $5f$ i-AlPdMn wide terrace after mean plane subtraction and drift correction.. Bottom: Fourier transform of the image (intensity in logarithmic scale). The radius of the inner circle of intense spots is $\pi / 0.62$ nm$^{-1}$.**

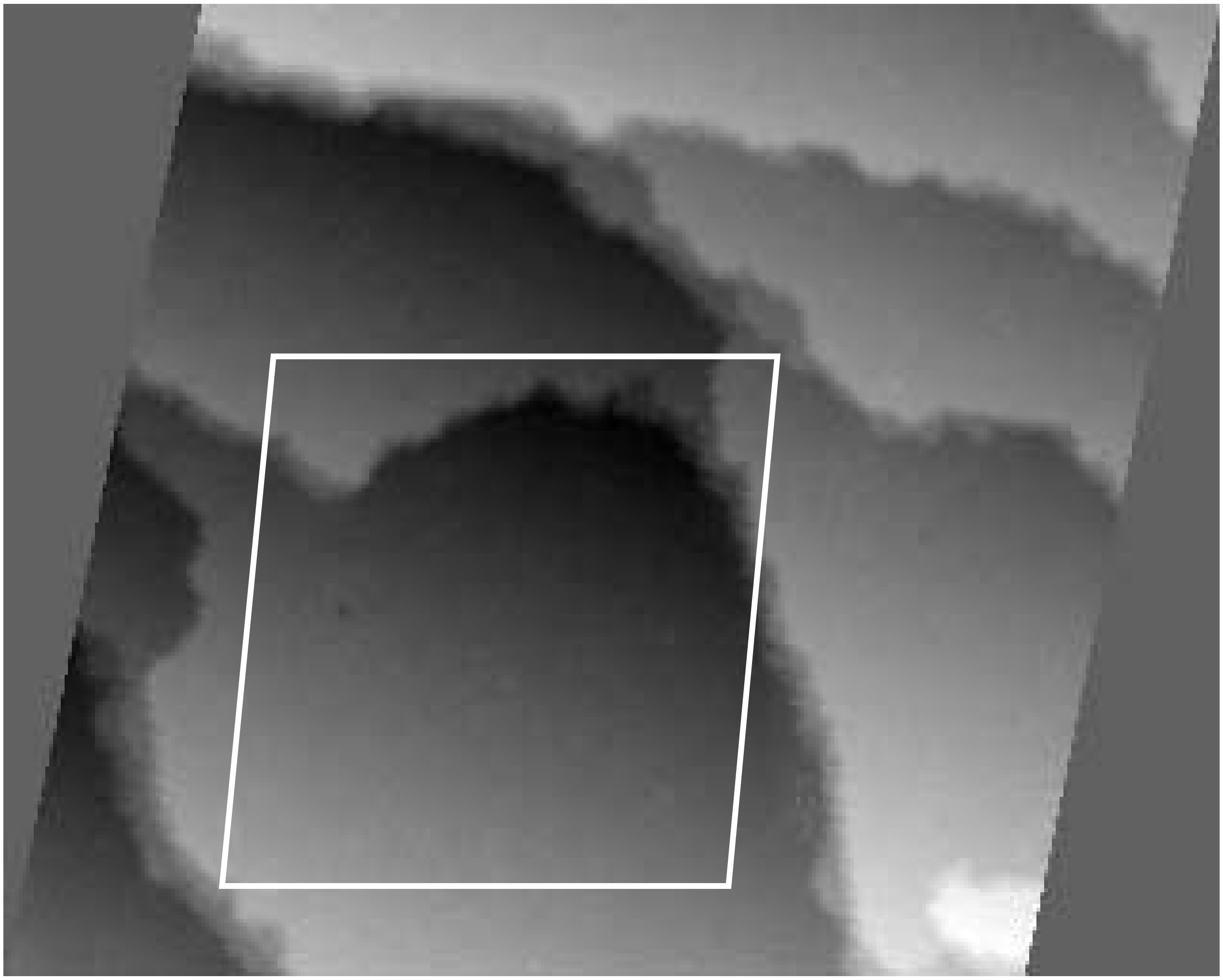

**Figure 19: 95.5 × 91.6 nm$^2$ STM image of the $5f$ surface of i-AlPdMn after mean plane subtraction and drift correction. ($V_t$ = -0.35 V, $I_t$ = 2.5 nA). STM resolution: 0.358 nm/pixel. The white bordered zone gives the location of the image of Figure 21.**

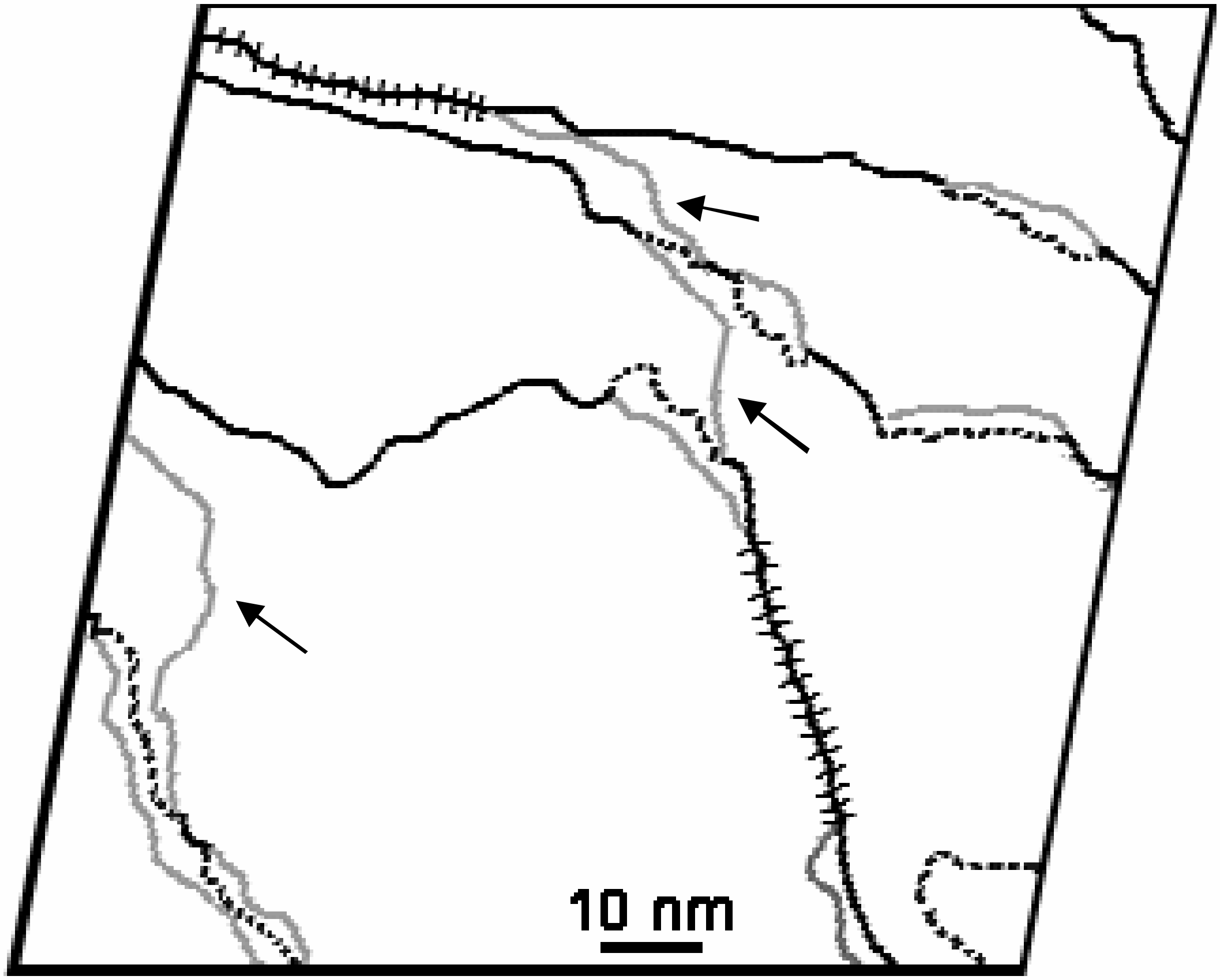

**Figure 20: Step heights in the image of Figure 19. (——) $(1+\tau)\,\delta_5$ = 0.68 nm; (grey line) $\tau\,\delta_5$, (- - - -) $\delta_5$, (++++) $(1+2\tau)\,\delta_5$. The arrows designate $\tau\,\delta_5$ steps leaving one bunch to join another. It is the signature of step bunching antiphase boundaries.**

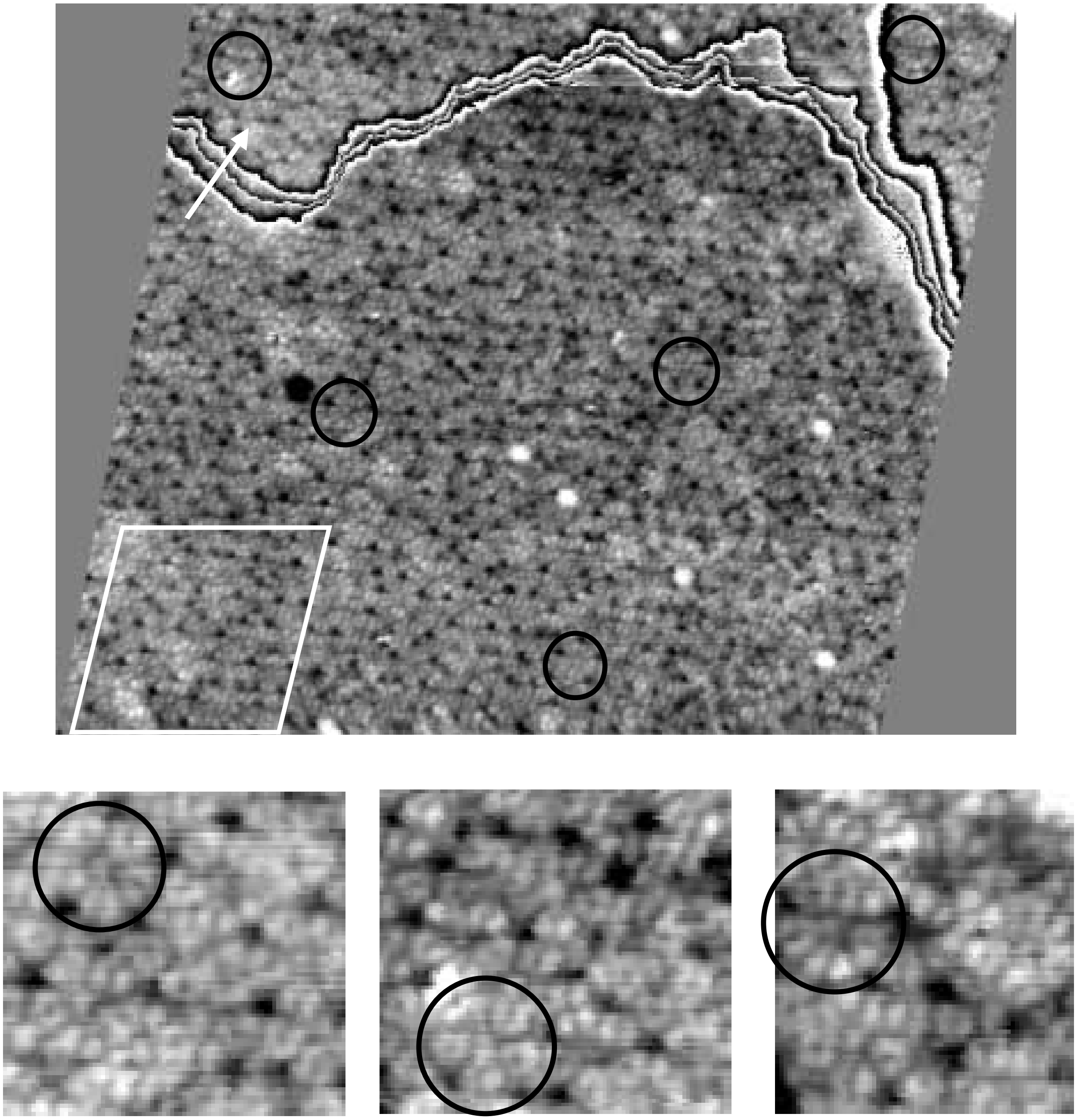

Lower Upper left Upper right

**Figure 21: Zoom (47.8 x 42.5 nm$^2$) of the zone indicated on Figure 19 after subtraction of the mean plane for each terrace and drift correction. STM resolution: 0.093 nm/pixel, $V_t$ = -0.45 V, $I_t$ = 2.5 nA. Along the white upper left arrow (bottom to top) the $(1+\tau)\,\delta_5$ height difference between the wide terraces results from the bunching of 3 separable steps of heights: $\delta_5$, $(\tau-1)\,\delta_5$ and $\delta_5$. Black circles in the main image and**

**in the 3 zoomed images of the lower, upper left and upper right terraces indicates the same pattern (filled flower) with the very same orientation indicating that wide terraces are isomorphic configurations.**

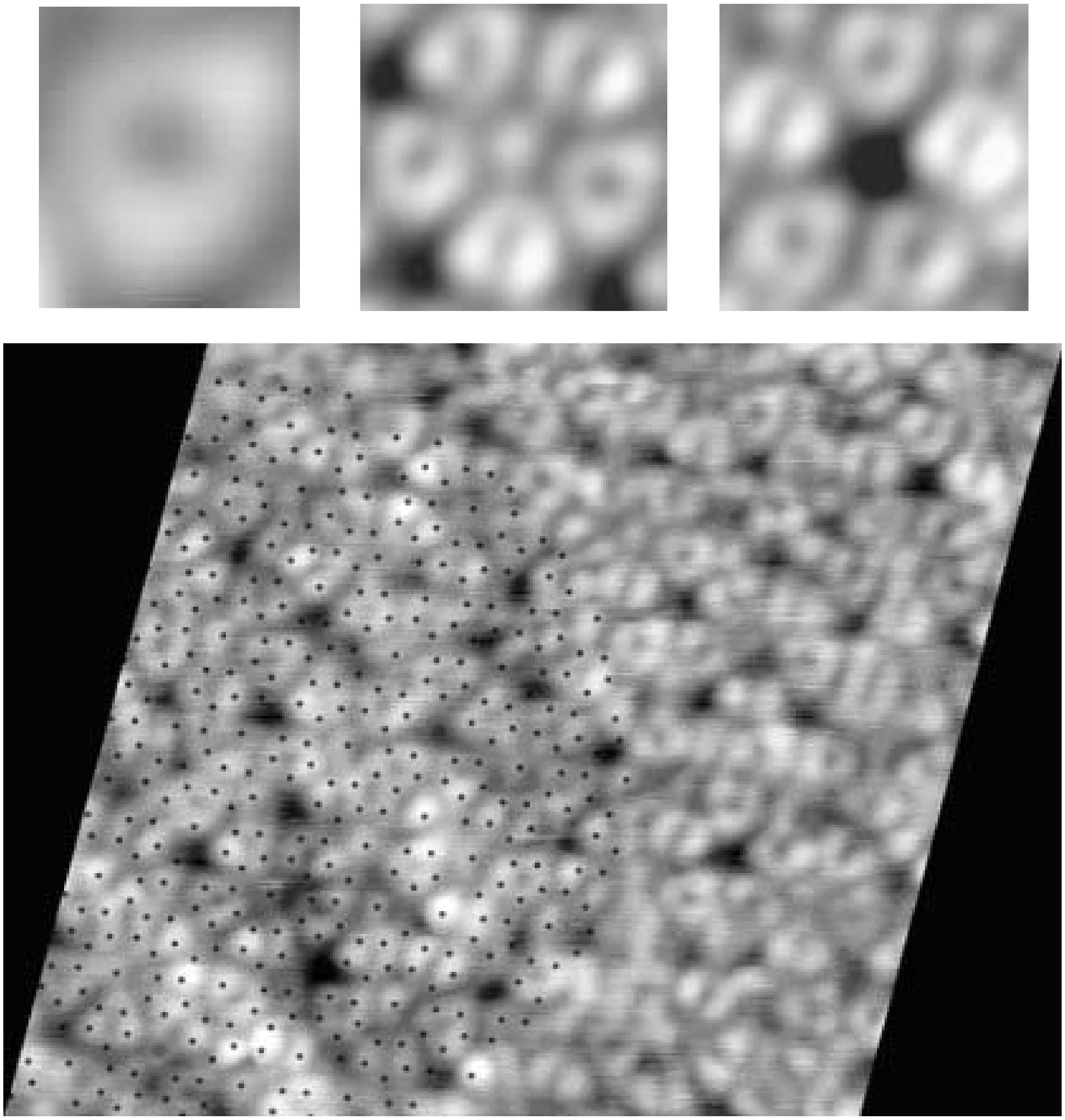

**Figure 22: Top (left to right): donut (0.87 x 0.75 nm$^2$) and filled and empty pentagonal flowers (2.2 x 2.2 nm$^2$) that are well recognizable patterns in high resolution images. Bottom: Zoom (11.9 x 11.6 nm$^2$) within the white bordered area indicated in Figure 21. Superimposed dark dots are in plane atoms issued from a cut of the $n$-AS at $(z, \delta z_{\perp}) = (-\tau, -\tau)\,\delta_5$ in almost good agreement with the STM image. STM resolution: 0.023 nm/pixel, $V_t$ = -0.45 V, $I_t$ = 2.5 nA.**

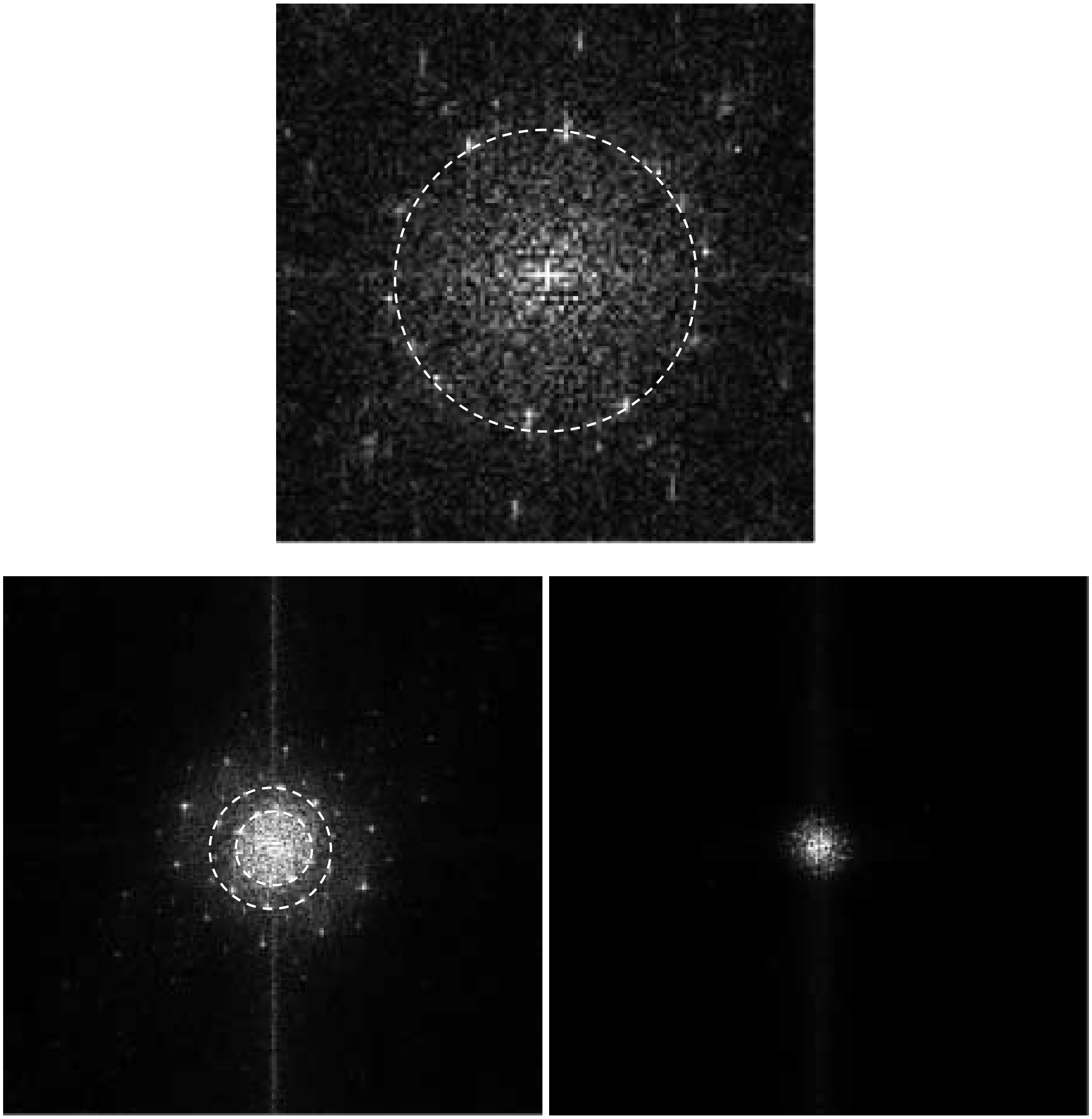

**Figure 23: Fourier transform of selected areas of Figure 19 (top), Figure 21 (bottom left) and Figure 22 (bottom right) within the same lower terrace. Scale : Inner circle $\pi / 0.6$ nm$^{-1}$. Intense $5f$ peaks are only present within the $\pi$ / 0.6 nm$^{-1}$- $\pi$ / 0.14 nm$^{-1}$ range, the typical donuts and flowers size.**

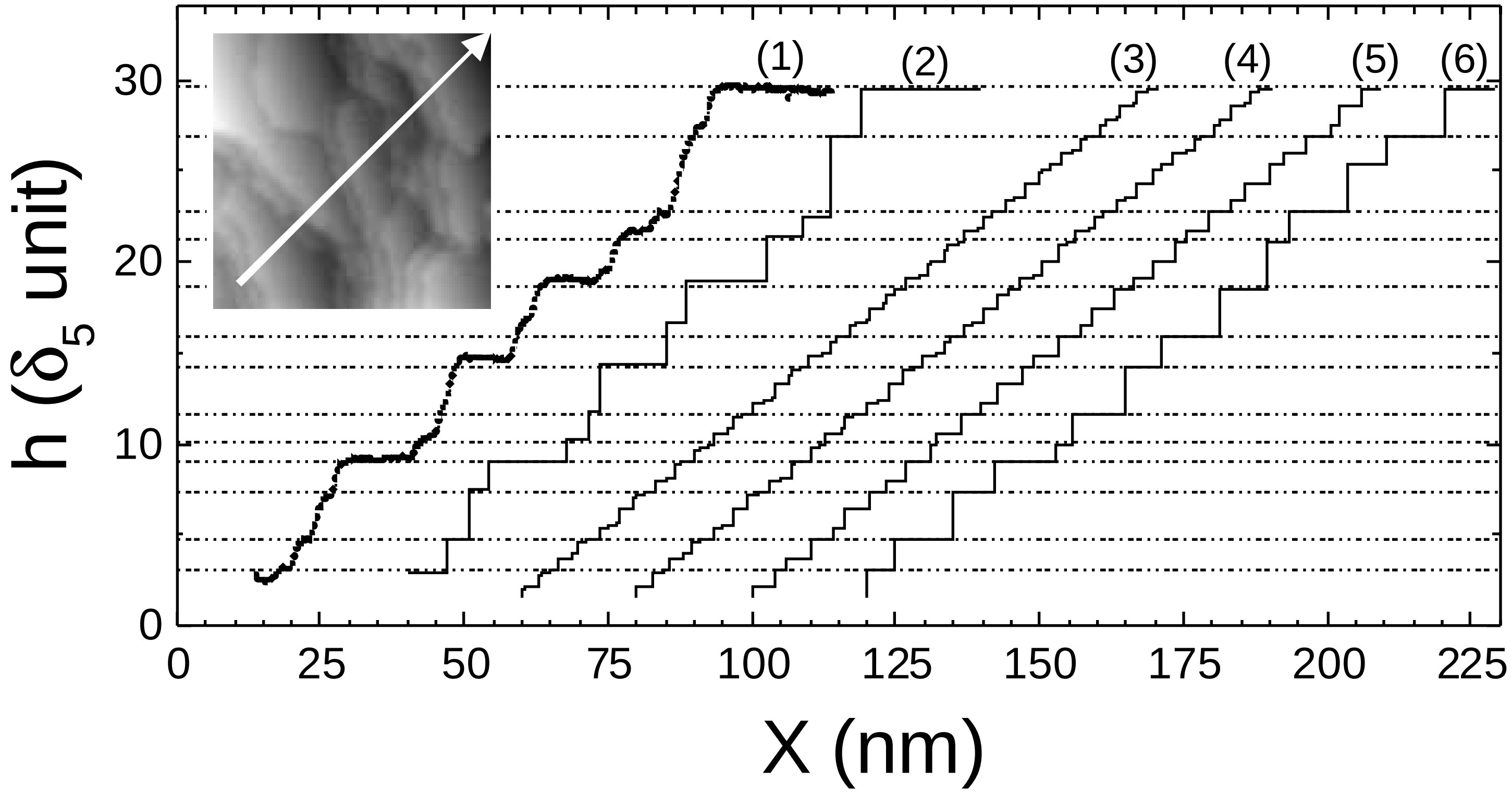

**Figure 24: Inset: STM image of a vicinal area. (1) Height profile along the diagonal of the STM image. (2) Steps and main terraces sequence in the STM image. Step-terrace sequences for the $5f$ orientation according to the polyhedral model (cut level: -0.315 $\lfloor E^6\ unit \rfloor$): (3) all AS are considered, terrace lengths are proportional to the in-plane density. (4) density selection: $bc$-AS and external slices of $n$-AS and $n'$-AS are rejected. (5) selection of dense Al-rich $n$-AS (chemical selectivity). (6) dense $n$-AS with underlying dense $n'$-AS at the cut level.**

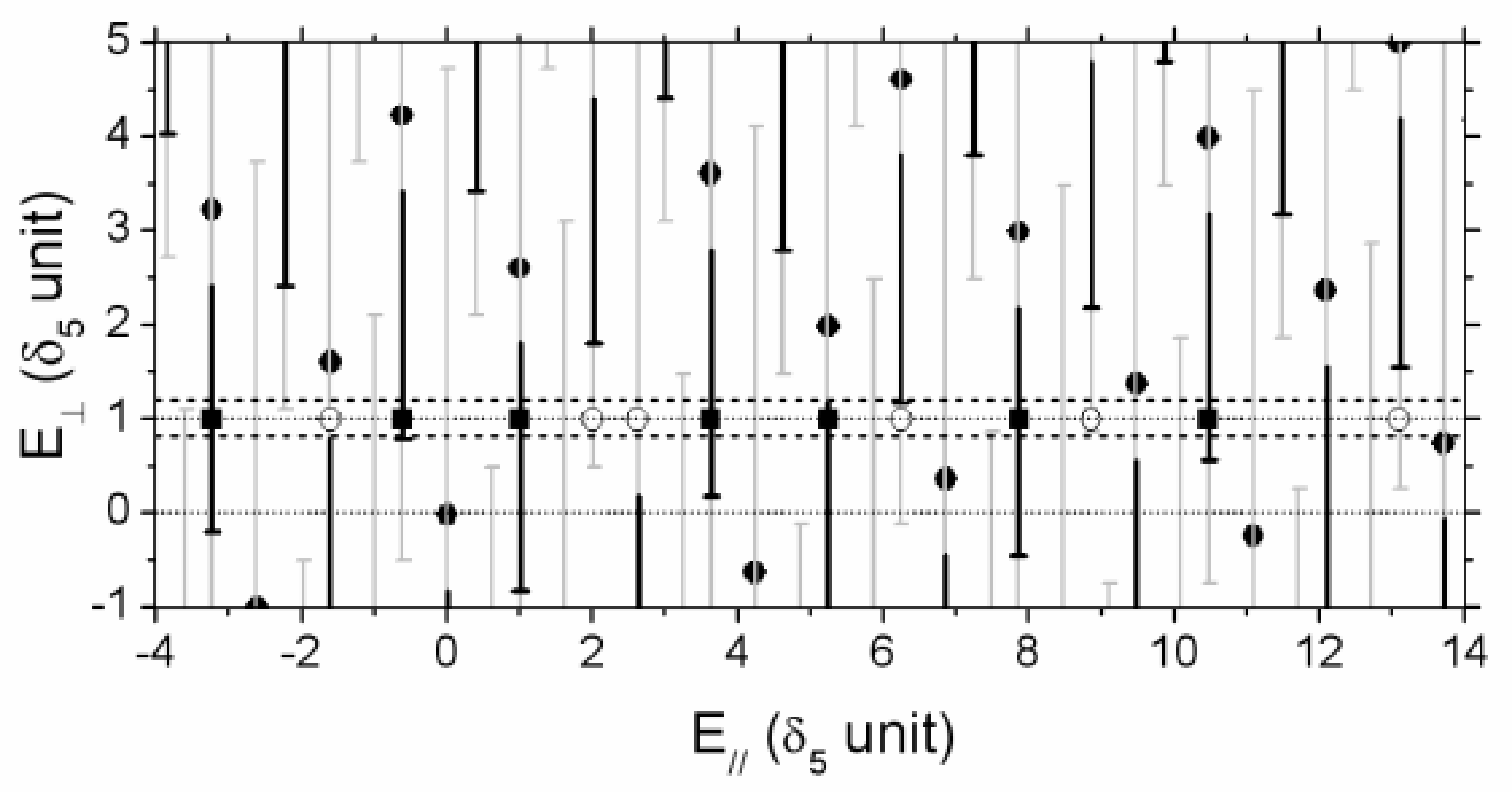

**Figure 25: Attribution of observed terraces in the image of Figure 19 to atomic surfaces of the polyhedral model. (■) position of wide terraces, (○) of intermediate terraces. Only $n$-AS (chemical selectivity) are drawn in $\Delta_Z(5f)$. Black part of atomic surfaces are the restriction of $n$-AS satisfying the 3 conditions: i) rejection of the outer slices by density selection, ii) rejection of the central decagonal part (no $10f$ pattern within wide terraces) and iii) presence of an underlying dense plane ($n'$-AS). Dashed lines delimitates the cut level range [0.809, 1.191] ($\delta_5$ unit) reproducing the step height sequence. All wide terraces results from a cut of $n$-AS within the 2^nd^-4^th^ lower slices. Other narrow terraces results also from a cut of $n$-AS but outside the selected part (in black).**

[24] J. Ledieu, R. McGrath, R.D. Diehl, T.A. Lograsso, D.W. Delaney, Z. Papadopolos, G. Kasner, Surf. Sci. Lett. **492**, L729 (2001).

[25] Z. Papadopolos, G. Kasner, J. Ledieu, E. J. Cox, N. V. Richardson, Q. Chen, R. D. Diehl, T. A. Lograsso, A. R. Ross and R. McGrath, Phys. Rev. B **66**, 184207 (2002).

[26] R. McGrath, J. Ledieu and R.D. Diehl, Progress in Surface Science **75**, 131 (2004).

[27] H. R. Sharma,V. Fournée, M. Shimoda, A. R. Ross, T. A. Lograsso, A. P. Tsai and A. Yamamoto, Phys. Rev. Lett. **93**, 165502 (2004).

[28] Quasicrystals, Editor P.A. Thiel, Progress in Surface Science **75**, (2004).

[29] D. Gratias, A. Katz and M. Quiquandon, J. Phys.: Condens. Matter **7**, 9101 (1995).

[30] As an alternative to the cut algorithm, it is equivalent to consider the cut and projection method where the real structure is defined by the projection in $E_{//}$ of all nodes of $\Lambda$ which projection in $E_{\perp}$ fall within an acceptance window. The size in $E_{\perp}$ of ASs and of acceptance windows are the same.

[31] Parity of a node is the parity of the sum of its 6 coordinates.

---

[32] F. Hippert, M. Audier, J. J. Préjean, A. Sulpice, E. Lhotel, V. Simonet, and Y. Calvayrac, Phys. Rev. B **68**, 134402 (2003).

[33] Beyond this simple (in $E_\perp$) model, a small void central triacontahedron can be inserted at the center of the $n$-AS introducing in that way a vacancy at the center of M and M' clusters making them closer to the original Mackay clusters (A.L. MacKay, Acta. Cryst. **15**, 916 (1962)). The atoms so removed (only $(-3+2\tau)/(11+14\tau)$=0.7% of atoms) from the $n$-AS center can be distributed at the periphery of $n$-AS or $n'$-AS to preserve the average atomic concentration (to be published).

[34] K. J. Franke, H. R. Sharma, W. Theis, P. Gille, Ph. Ebert and K. H. Rieder, Phys Rev. Lett. **89**, 156104 (2002).

[35] H. R. Sharma, K. J. Franke, W. Theis, P. Gille, Ph. Ebert and K. H. Rieder, Phys. Rev. B **68**, 054205 (2003).

[36] H.R. Sharma, W. Theis, P. Gille and K.H. Rieder, Surf. Sci. **511**, 387 (2002), H. R. Sharma, K. J. Franke, W. Theis, A. Riemann, S. Fölsch, K. H. Rieder and P. Gille, Surf. Sci. **561**, 121 (2004).

[37] B. Bolliger, M. Erbudak, D. D. Vvedensky, M. Zurkirch and A. R. Kortan, Phys. Rev. Lett. **80**, 5369 (1998).

---

[46] The cubic unit cell with lattice constant $2a$ of the $F(2)$ lattice contains 32 even nodes and the atomic density generated by one AS is: $\rho_{at} = 32\, V_{AS} / 2^6 \left[ at / \left(E^6\ unit\right)^3 \right]$. The total atomic density for AlPdMn is 64.40 at/nm$^3$. (60.24 at/nm$^3$ for fcc Al).

[47] G. Armand , D. Gorse, J. Lapujoulade and J. R. Manson, Europhys Lett. **3** (10), 1113 (1987).

[48] J. Lapujoulade, Surface Science **134**, L529 (1983).